\documentclass[11pt,letterpaper]{article}

\usepackage{times}
\usepackage{relsize}
\usepackage{microtype}
\usepackage{xspace}
\usepackage[table,xcdraw,dvipsnames]{xcolor}
\usepackage{color,colortbl}
\usepackage{subfigure}
\usepackage{alltt}
\usepackage{url}
\usepackage{comment}
\usepackage{graphicx}
\usepackage{relsize}
\usepackage{paralist}
\usepackage{todonotes}
\usepackage{wrapfig}
\usepackage{titlesec}
\usepackage{epsfig}
\usepackage{geometry}
\usepackage{tikz}
\usepackage[normalem]{ulem}
\usepackage{booktabs}
\usepackage{wrapfig}
\usepackage{multirow}
\usepackage{pifont}
\usepackage{wrapfig}
\usepackage{soul}
\usepackage{enumitem}
\usepackage{tcolorbox}
\usepackage{pdfpages}
\usepackage{fancyhdr}
\usepackage{longtable}
\usepackage{makecell}
\usepackage{lettrine}
\usepackage{listings}
\usepackage{longtable}
\usepackage{tcolorbox}
\usepackage[numbers]{natbib}
\usepackage{wrapfig}

\AtBeginDocument{
  \definecolor{pdfurlcolor}{HTML}{831c15}
  \definecolor{pdfcitecolor}{rgb}{0,0.6,0}
  \definecolor{pdflinkcolor}{HTML}{831c15}
  \definecolor{light}{gray}{.85}
  \definecolor{vlight}{gray}{.95}
}
\usepackage[colorlinks=true,citecolor=pdfcitecolor,urlcolor=pdfurlcolor,linkcolor=pdflinkcolor,pdfborder={0 0 0}]{hyperref}

\titlespacing{\section}{0pc}{1pc}{0.5pc}
\titlespacing{\subsection}{0pc}{1pc}{0.5pc}
\titlespacing{\subsubsection}{0pc}{0.5pc}{0.5pc}
\titleformat*{\section}{\Large\bfseries}
\titleformat*{\subsection}{\large\bfseries}
\titleformat*{\subsubsection}{\normalsize\bfseries}

\lstdefinestyle{wcsStyle}{
  columns=fullflexible,
  tabsize=2,
  showspaces=false,
  showstringspaces=false,
  aboveskip=0em,
  belowskip=0em,
}

\definecolor{mypurple}{HTML}{f4ebfd}
\definecolor{myorange}{RGB}{255, 243, 219}
\definecolor{myorange2}{RGB}{255, 170, 23}
\definecolor{ruler}{HTML}{831c15}

\newcommand{\recommendations}[2][inline]{
\begin{tcolorbox}[width=\linewidth,colback={mypurple},colframe={ruler}] \textbf{\color{ruler} Recommendations:} #2 \xspace \end{tcolorbox}}

\definecolor{Gray}{gray}{0.9}

\newcommand{\pp}[1]{\medskip \noindent \textbf{\emph{#1.}}\xspace}

\let\oldheadrule\headrule
\renewcommand{\headrule}{\color{ruler}\oldheadrule}

\DeclareUrlCommand\url{\color{ruler}}

\begin{document}

\includepdf[pages=-]{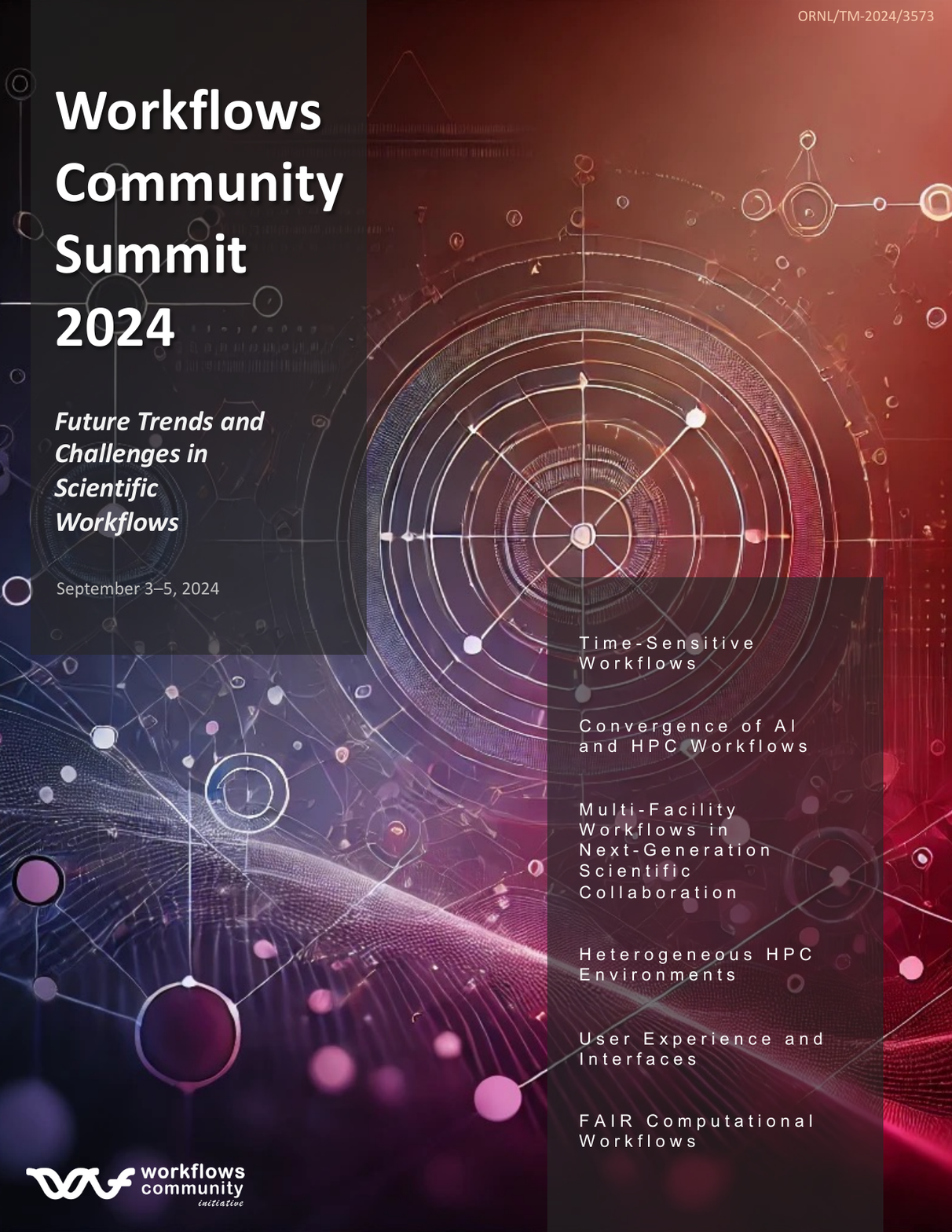}


\pagestyle{fancy}
\fancyhf{}
\rhead{
  \includegraphics[height=11pt]{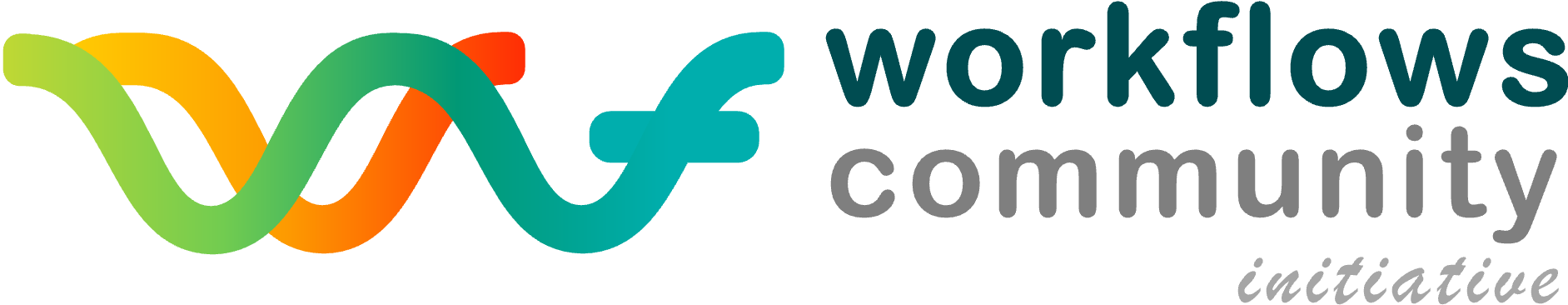}
}
\lhead{\color{gray}\smaller WORKFLOWS COMMUNITY SUMMIT 2024}
\rfoot{\thepage}


\begin{table}[!ht]
\centering
\smaller
\begin{tabular}{p{16cm}}
    \textbf{Disclaimer.}
    This research used resources of the Oak Ridge Leadership Computing Facility at the Oak Ridge National Laboratory, which is supported by the Office of Science of the U.S. Department of Energy under Contract No. DE-AC05-00OR22725.
    \\ 
    \vspace{-0.25em}
    This report was prepared as an account of work sponsored by agencies of the United States Government. Neither the United States Government nor any agency thereof, nor any of their employees, makes any warranty, express or implied, or assumes any legal liability or responsibility for the accuracy, completeness, or usefulness of any information, apparatus, product, or process disclosed, or represents that its use would not infringe privately owned rights. Reference herein to any specific commercial product, process, or service by trade name, trademark, manufacturer, or otherwise, does not necessarily constitute or imply its endorsement, recommendation, or favoring by the United States Government or any agency thereof. The views and opinions of authors expressed herein do not necessarily state or reflect those of the United States Government or any agency thereof.
    \\
    \vspace{-0.25em}
    \textbf{License.}
    This report is made available under a Creative Commons Attribution 4.0 International Public license ({\small \url{https://creativecommons.org/licenses/by/4.0}}).
\end{tabular}
\end{table}

\vspace{-0.5em}
\begin{table}[!ht]
\centering
\smaller
\begin{tabular}{p{16cm}}
    \textbf{\small Preferred citation} 
    \\
    R. Ferreira da Silva, D. Bard, K. Chard, S. DeWitt, I. T. Foster, T. Gibbs, C. Goble, W. Godoy, J. Gustafsson, U.-U. Haus
% ...
    ``\emph{Workflows Community Summit 2024: Future Trends and Challenges in Scientific Workflows}", Technical Report, ORNL/TM-2024/3573, October 2024, DOI: 10.5281/zenodo.13844759.
    \\
    \rowcolor[HTML]{F7F7F7}
    \lstset{basicstyle=\scriptsize,style=wcsStyle}
    \begin{lstlisting}
@techreport{wcs2024,
  author      = {Ferreira da Silva, Rafael and Bard, Deborah and Chard, Kyle and DeWitt, Shaun and Foster, Ian T. and Gibbs, Tom and Goble, 
            Carole and Godoy, William and Gustafsson, Johan and Haus, Utz-Uwe and Hudson, Stephen and Jha, Shantenu and Los, 
            Laila and Paine, Drew and Suter, Frederic and Ward, Logan and Wilkinson, Sean and Amaris, Marcos and Babuji, Yadu and 
            Bader, Jonathan and Balin, Riccardo and Balouek, Daniel and Beecroft, Sarah and Belhajjame, Khalid and Bhattarai, Rajat and 
            Brewer, Wes and Brunk, Paul and Caino-Lores, Silvina and Casanova, Henri and Cassol, Daniela and Coleman, Jared and 
            Coleman, Taina and Colonnelli, Iacopo and Da Silva, Anderson Andrei and de Oliveira, Daniel and Elahi, Pascal and 
            Elfaramawy, Nour and Elwasif, Wael and Etz, Brian and Fahringer, Thomas and Ferreira, Wesley and Filgueira, Rosa and Fosso 
            Tande, Jacob and Gadelha, Luiz and Gallo, Andy and Garijo, Daniel and Georgiou, Yiannis and Gritsch, Philipp and Grubel, 
            Patricia and Gueroudji, Amal and Guilloteau, Quentin and Hamalainen, Carlo and Hong Enriquez, Rolando and Huet, Lauren 
            and Hunter Kesling, Kevin and Iborra, Paula and Jahangiri, Shiva and Janssen, Jan and Jordan, Joe and Kanwal, Sehrish and 
            Kunstmann, Liliane and Lehmann, Fabian and Leser, Ulf and Li, Chen and Liu, Peini and Luettgau, Jakob and Lupat, Richard 
            and Fernandez, Jose M. and Maheshwari, Ketan and Malik, Tanu and Marquez, Jack and Matsuda, Motohiko and Medic, 
            Doriana and Mohammadi, Somayeh and Mulone, Alberto and Navarro, John-Luke and Ng, Kin Wai and Noelp, Klaus and 
            P. Kinoshita, Bruno and Prout, Ryan and R. Crusoe, Michael and Ristov, Sashko and Robila, Stefan and Rosendo, Daniel and 
            Rowell, Billy and Rybicki, Jedrzej and Sanchez, Hector and Saurabh, Nishant and Saurav, Sumit Kumar and Scogland, Tom and 
            Senanayake, Dinindu and Shin, Woong and Sirvent, Raul and Skluzacek, Tyler and Sly-Delgado, Barry and Soiland-Reyes, 
            Stian and Souza, Abel and Souza, Renan and Talia, Domenico and Tallent, Nathan and Thamsen, Lauritz and Titov, Mikhail and 
            Tovar, Ben and Vahi, Karan and Vardar-Irrgang, Eric and Vartina, Edite and Wang, Yuandou and Wouters, Merridee and Yu, Qi 
            and Al Bkhetan, Ziad and Zulfiqar, Mahnoor},
  title       = {{Workflows Community Summit 2024: Future Trends and  Challenges in Scientific Workflows}},
  year        = {2024},
  publisher   = {Zenodo},
  number      = {ORNL/TM-2024/3573},
  doi         = {10.5281/zenodo.13844759},
  institution = {Oak Ridge National Laboratory}
}
    \end{lstlisting}
    \\
\end{tabular}
\end{table}

\newpage

\tableofcontents

\newpage


\cleardoublepage\phantomsection\addcontentsline{toc}{section}{Executive Summary}
\section*{Executive Summary}
\label{sec:summary}

\begin{wrapfigure}{R}{0.54\textwidth} 
    \vspace{-45pt}
    \begin{tcolorbox}[width=\linewidth,colback={mypurple},colframe={ruler}] 
    \small
    {\color{ruler} \bf KEY RECOMMENDATIONS}
    \\
    \textbf{Time Sensitive:} Develop standardized cross-facility authentication and data protocols to enable secure, near real-time data movement and analysis across distributed research infrastructures.
    \\
    \textbf{AI-HPC:} Invest in R\&D of AI-driven workflow optimization techniques to automate the design and execution of complex HPC workflows.
    \\
    \textbf{Multi-Facility:} Implement distributed-by-design workflow models to create workflows that inherently account for multi-facility execution, incorporating environment descriptions and constraints.
    \\
    \textbf{Heterogeneous Environments:} Develop a community-driven suite of diverse workflow benchmarks to guide system evaluation, procurement, and performance optimization across heterogeneous HPC environments.
    \\
    \textbf{UX and Interfaces:} Formulate and implement a set of UX principles specifically tailored for scientific workflows to guide the design of more intuitive, user-friendly interfaces across multiple workflow systems.
    \\
    \textbf{FAIR:} Create a FAIR workflow maturity model and associated metrics to guide the implementation of FAIR principles and provide a standardized way to assess and improve workflow FAIRness.
    \end{tcolorbox}
    \vspace{-20pt}
\end{wrapfigure}

The Workflows Community Summit convened 111 participants from 18 countries to explore emerging trends and challenges in scientific workflows. It focused on six key areas that are shaping the future of computational research: time-sensitive workflows, the convergence of AI and HPC, multi-facility workflows, heterogeneous HPC environments, user experience and interfaces, and FAIR computational workflows.

The integration of AI and exascale computing has revolutionized scientific workflows, enabling higher-fidelity models and more complex, time-sensitive processes. However, this advancement introduces significant challenges in managing heterogeneous environments and multi-facility data dependencies. To address these issues, innovative approaches like trust bundles and advanced data fabric solutions are being developed. Simultaneously, the rise of large language models is pushing computational demands to zettaflop scales, driving the development of modular, adaptable systems and cloud-service models. These new paradigms aim to optimize resource utilization, enhance scalability, and ensure reproducibility across diverse computing infrastructures.

As data volumes and velocity continue to grow, there is an increasing demand for workflows that span multiple facilities. This shift presents challenges in data movement, curation of large-scale data, and overcoming institutional silos. The increasing prevalence of diverse hardware architectures, including specialized AI accelerators, requires a focus on integrating workflow considerations into early system design stages and developing standardized resource and data management~tools. The summit also emphasized the critical importance of user experience in workflow systems, with on-going efforts to simplify workflow usage, improve graphical representations, and adapt concepts from commercial software development to scientific applications. Additionally, ensuring Findable, Accessible, Interoperable, and Reusable (FAIR) workflows is crucial for enhancing collaboration and accelerating scientific discovery, but challenges remain in balancing standardization with innovation and applying FAIR principles to workflows with confidential components.

Key recommendations include developing standardized metrics for time-sensitive workflows, creating frameworks for cloud-HPC integration, implementing distributed-by-design workflow modeling, establishing multi-facility authentication protocols, and accelerating AI integration in HPC workflow management. The summit also emphasized the need for comprehensive workflow benchmarks, formulating workflow-specific UX principles, and creating a FAIR workflow maturity model. These recommendations underscore the importance of continued collaboration and innovation in addressing the complex challenges posed by the convergence of AI, HPC, and multi-facility research environments. 

\newpage

\section{Introduction}
\label{sec:introduction}

The landscape of scientific workflows is undergoing a profound transformation, driven by the convergence of high performance computing (HPC) and artificial intelligence (AI)~\cite{ejarque2022enabling, brewer2024ai}. Scientific workflows are evolving from being mere tools to becoming the new applications that drive forward the frontiers of science. This evolution is characterized by increasing complexity, data volume, and computational demands, thus necessitating robust, adaptable, and flexible support systems. Key trends that shape the future of scientific workflows include the deep integration of AI with HPC, the rise of multi-facility workflows spanning multiple research institutions, the shift towards data-driven workflow dynamics, the challenges posed by heterogeneous computing environments, and the pressing need for sustainable, energy-efficient practices~\cite{ferreiradasilva2024computer, badia2024integrating}. These developments present both exciting opportunities and significant challenges for the community, requiring innovative approaches to workflow design, management, and execution. As we enter the postexascale era, the demand for processing scientific data continues to grow, alongside steady enhancements in large-scale HPC capabilities. The integration of edge computing with traditional HPC resources is creating a more fluid and efficient computing continuum~\cite{antypas2021enabling}, further complicating the workflow landscape. Additionally, the need for near real-time analysis capabilities and improved workflow resilience is driving the development of more dynamic and flexible data management solutions. The Workflows Community Summit aims to address these emerging trends and challenges, fostering collaboration and knowledge exchange to shape the future of scientific workflows in an era of unprecedented technological advancement and environmental awareness.

\subsection{The Workflows Community Summit 2024}

\begin{figure}[!ht]
    \centering
    \includegraphics[width=.9\linewidth]{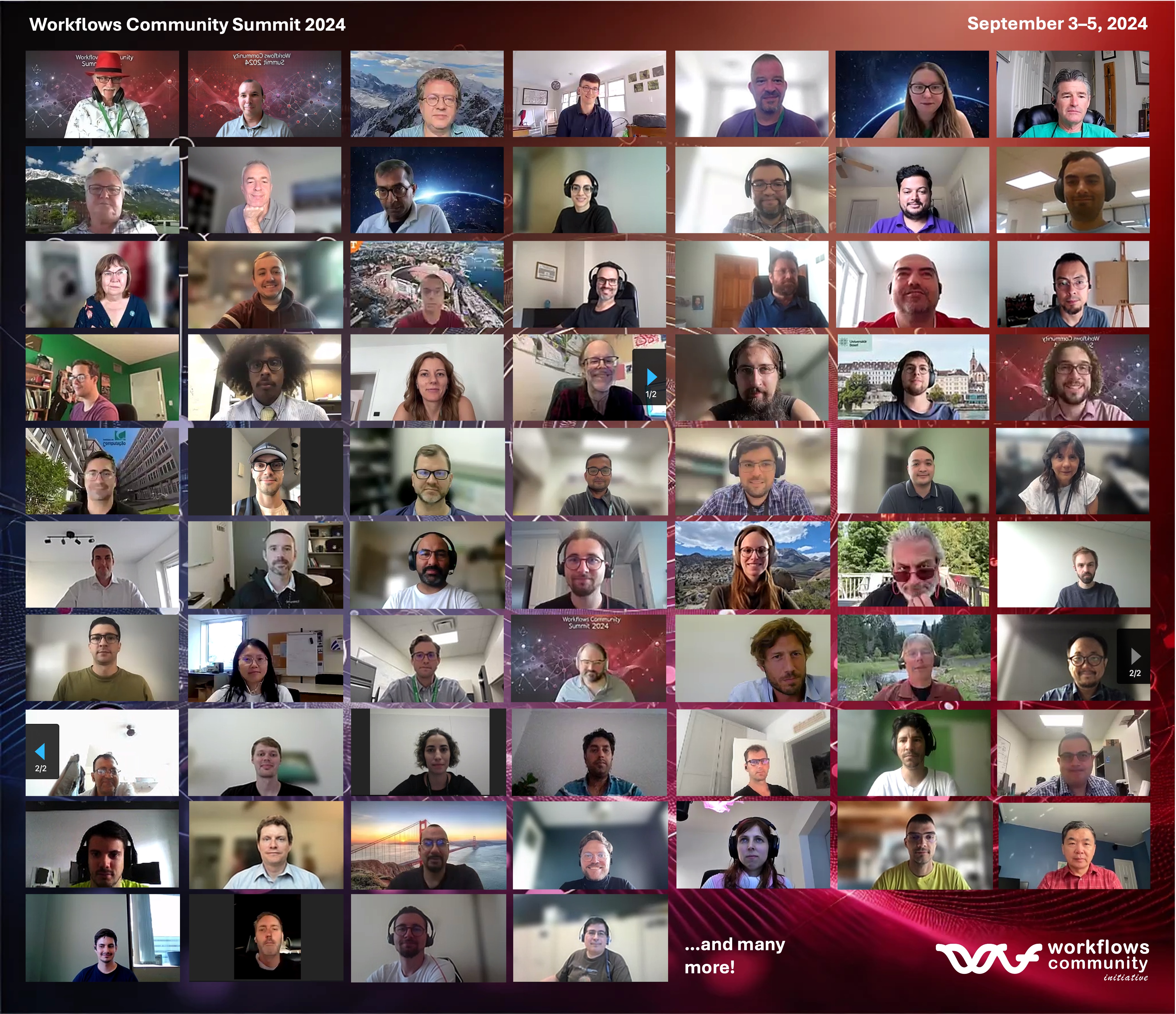} \\
    \vspace{-1em}
    \caption{Participants of the 2024 Workflows Community Summit (virtual, Sept. 3--5, 2024).}
    \label{fig:participants}
\end{figure}

This document presents the outcomes and insights from the 2024 international edition of the ``Workflows Community Summit," held from September 3--5, 2024~\cite{wcs-2024}. The three-day summit brought together a diverse group of \textbf{111 participants} (Figure~\ref{fig:participants}), representing a global community of researchers, developers, and industry professionals. The event expanded for the first time, adding an extra day with a timezone favorable for Oceania and Asia. Attendees represented \textbf{18 countries} across six continents, including Australia, Austria, Brazil, France, Germany, India, Italy, Japan, Latvia, New Zealand, Puerto Rico, Singapore, Spain, Sweden, Switzerland, The Netherlands, United Kingdom, and the United States. This international cohort comprised experts from various workflow management systems (WMSs), end-users, and industry representatives, fostering a rich environment for cross-cultural and interdisciplinary exchange. The summit's increasing global reach underscores the increasing importance of scientific workflows in addressing complex, large-scale research challenges that transcend national boundaries. In future years, the event will seek to expand its reach further, inviting contributions from workflow experts across domains and disciplines.

The summit was organized by key members of the Workflows Community Initiative (WCI)~\cite{wci}, a volunteer-driven organization dedicated to unifying the diverse workflows community. The WCI's mission is to provide valuable community resources and capabilities that empower scientists and workflow system developers. The initiative serves as a catalyst for community-wide efforts to address the grand challenges facing scientific workflows. By facilitating this summit, the WCI continues its commitment to advancing the state-of-the-art in workflow technologies and methodologies, ultimately accelerating scientific discovery and innovation across disciplines.

\subsection{Summit Structure and Activities}

Building on trends discussed in~\cite{ferreiradasilva2024computer}, the six cross-cutting summit's topics reflect emerging challenges and opportunities in the rapidly evolving landscape of scientific computing (Table~\ref{tab:topics}): (i)~``Convergence of AI and HPC Workflows" addresses the increasing synergy between AI systems and traditional HPC simulations; (ii)~``Multi-Facility Workflows in Next-Generation Scientific Collaboration" explores the complexities of orchestrating workflows across diverse research infrastructures; (iii)~``Time-Sensitive Workflows" focuses on near real-time data integration and processing challenges; (iv)~``Heterogeneous HPC Environments" tackles the need to navigate performance variability across diverse hardware configurations; (v)~``User Experience and Interfaces" recognizes the growing importance of accessibility and usability in complex scientific workflows; and (vi)~``FAIR Computational Workflows" emphasizes the need for findable, accessible, interoperable, and reusable workflow practices. Collectively, these topics address critical trends and challenges identified in the previous years' summits, aiming to shape the future of scientific workflows in the context of rapidly evolving HPC technologies and practices.

\begin{table}[!ht]
    \centering
    \smaller
    \setlength{\tabcolsep}{4pt}
    \begin{tabular}{p{7.2cm}p{8.7cm}}
        \toprule
        \textbf{Topic} & \textbf{Discussion Co-Leaders} \\
        \midrule
        \rowcolor[HTML]{F2F2F2}
        Topic 1: Time-Sensitive Workflows & Utz-Uwe Haus (HPE), Shaun DeWitt (UKAEA) \\
        Topic 2: Convergence of AI and HPC Workflows & Shantenu Jha (PPPL), Logan Ward (ANL) \\
        \rowcolor[HTML]{F2F2F2}
        Topic 3: Multi-Facility Workflows in Next-Generation Scientific Collaboration & Tom Gibbs (NVIDIA), Debbie Bard (LBNL) \\
        Topic 4: Heterogeneous HPC Environments & William Godoy (ORNL), Stephen Hudson (ANL) \\
        \rowcolor[HTML]{F2F2F2}
        Topic 5: User Experience and Interfaces & Drew Paine (LBNL), Laila Los (UFreiburg) \\
        Topic 6: FAIR Computational Workflows & Carole Goble (UManchester), Sean Wilkinson (ORNL) \\
        \rowcolor[HTML]{F2F2F2}
        Spotlight Discussion: What makes a workflow good? & Johan Gustafsson (Australian BioCommons, University of Melbourne) \\
        \bottomrule
    \end{tabular}
    \vspace{-1.2em}
    \caption{Workflows Community Summit topics and discussion co-leaders.}
    \vspace{-1.2em}
    \label{tab:topics}
\end{table}

For each of the technical topics listed in Table~\ref{tab:topics}, the co-leaders delivered concise 10-minute lightning talks in plenary sessions, followed by in-depth discussions in targeted breakout groups. These sessions had dual objectives: (i)~to evaluate previously identified challenges and their proposed solutions, and (ii)~to pinpoint critical gaps and formulate potential short- and long-term strategies for supporting emerging and novel workflow applications in light of the rapidly evolving computing continuum paradigm. Day~3 of the event featured a session of lightning talks highlighting workflow-related activities from Oceania and Asia.

\vspace{-2pt}
\begin{tcolorbox}[width=\linewidth,colback={mypurple},colframe={ruler}]
All presentations and videos can be found in the summit website (\url{https://workflows.community/summits/2024}), and videos can be watched from WCI's YouTube channel (\url{https://www.youtube.com/watch?v=STy5HwEVj8k&list=PLAtmuqHExRvOzPOvJfSH8PwjbVJFyulCy}).
\end{tcolorbox}
\vspace{-8pt}

\pp{Keynote Speaker}
The summit was further enriched by an invited keynote address from \textbf{\emph{Ian T. Foster}} (Argonne National Laboratory / University of Chicago), a distinguished leader in Scientific Computing, who spoke on ``\textbf{\emph{AI-mediated Scientific Workflows}}." Dr. Foster presented a vision of accelerated scientific discovery through the integration of AI methods into various stages of the scientific process. He discussed the development of AI-driven ``embodied agents" capable of performing tasks such as literature review, hypothesis generation, and experimental design. Foster highlighted a specific example in antimicrobial peptide discovery, where AI agents can interact with databases, run simulations, and even control robotic laboratories. He emphasized the need for specialized language models designed for scientific applications, and the importance of creating persistent, self-improving AI systems. Foster also addressed the computational challenges of these AI-mediated workflows, including the need for secure, reliable, and scalable infrastructure to support long-running, stateful processes across diverse computing environments.

\pp{Spotlight Discussion}
The spotlight discussion ``\emph{\textbf{What makes a workflow good?}}", led by Johan Gustafsson from Australian BioCommons \cite{francis_australian_2024}, explored the intricacies of workflow quality in scientific computing. Participants addressed the tension between research outcomes and the demand for accessible, reproducible workflows. Key attributes of exemplary workflows include containerization for stability, rigorous validation, comprehensive documentation, and ongoing support. Standard software practices, especially unit testing, and the role of Research Software Engineers in developing and maintaining high-quality workflows were emphasized for reliability. Challenges discussed included balancing user-friendliness with preventing misuse, and managing diverse expertise across the workflow lifecycle. A ``good" workflow was defined as robust, purpose-driven, environment-agnostic, and well-documented. This holistic approach to workflow quality reflects the evolving scientific computing landscape, demonstrating a shared commitment to sustainable, reproducible methodologies that advance scientific discovery.

\section{A Look Back at Workflows Community Summits}
\label{sec:overview}

The workflows community took a significant step forward with the publication of ``A Community Roadmap for Scientific Workflows Research and Development"~\cite{ferreiradasilva2021works} in 2021. This pioneering paper, resulting from the first series of Workflows Community Summits, presented a consolidated view of the state of the art, challenges, and potential future directions in the field of scientific workflows. The roadmap identified six key themes: FAIR computational workflows, AI workflows, exascale challenges, APIs and interoperability, training and education, and building a workflows community. For each theme, the roadmap outlined specific challenges and proposed community activities to address them. Additionally, it provided a set of technical milestones across three main thrusts: defining common workflow patterns and benchmarks, identifying paths toward interoperability, and improving workflow systems' interface with HPC software and hardware stacks.

Building on this foundation, the Workflows Community Summit 2022 further advanced the discourse on scientific workflows. The summit report~\cite{wcs2022} expanded on the previous themes and introduced new focus areas: specifications, standards, and APIs; AI workflows; high-performance data management and in situ workflows; HPC and quantum workflows; FAIR computational workflows; and workflows for continuum and cross-facility computing. Each topic was thoroughly discussed, with participants identifying current challenges and proposing both short- and long-term solutions. The summit highlighted the evolving landscape of scientific computing and the need for WMSs to adapt to emerging paradigms such as edge-to-HPC computational workflows and the integration of quantum computing with classical HPC systems.

Most recently, the paper ``Frontiers in Scientific Workflows: Pervasive Integration with High-Performance Computing"~\cite{ferreiradasilva2024computer} provides a forward-looking perspective on the field by identifying five key trends shaping the future of workflows: increased synergy between AI and HPC workflows, the rise of cross-facility workflows, data-driven HPC workflow dynamics, managing performance variability in heterogeneous HPC environments, and the push for sustainable practices in HPC workflow design. For each trend, the paper predicts future developments: AI systems for science trained on HPC workflow data, global interoperability standards for multi-facility collaboration, dynamic near real-time workflow control planes, software solutions for consistent performance across heterogeneous hardware, and AI-driven optimization of HPC workflows for energy efficiency and reduced environmental impact. It emphasizes the field's rapid evolution and scientific workflows' key role in advancing cross-disciplinary research.


\section{Time-Sensitive Workflows}
\label{sec:time-sensitive}

Time-sensitive workflows have been integral to a diverse range of scientific and technological domains, from autonomous vehicles and aircraft control systems to cutting-edge research in beamline-based materials science, astronomy and astrophysics, observational science, and experimental fusion~\cite{brown2022workflows}. These workflows are characterized by their urgent nature, requiring real-time or near real-time responses across multiple facilities or resources to enable timely decision-making, experiment steering, and data analysis~\cite{brown2023integrated}.

\pp{Emerging Paradigms and Challenges}
Since the 2022 summit, significant changes have occurred in this field, driven by advancements in AI, machine learning, and exascale computing. The rise of AI and machine learning has led to the increased use of surrogate models, replacing traditional reduced-order models due to their speed and efficiency in time-critical situations. This shift has been particularly noticeable in autonomous vehicles and other systems requiring rapid decision-making. Exascale computing has enabled larger models with higher fidelity and smaller time steps, pushing the boundaries of what is possible in computational science. Additionally, the volume of data generated by exascale systems, such as the petabyte-per-day output from projects like Destination Earth~\cite{wedi2022destination}, poses significant data movement and management challenges. The integration of digital twins, the need for federation across facilities, and the growing importance of provenance in regulated environments have added new layers of complexity. These advancements have brought forth challenges in managing heterogeneous computing environments, dealing with cross-facility data dependencies, and addressing on-demand access and authentication issues in time-critical scenarios.

\pp{Recent Advances and Emerging Solutions}
To address these challenges, several state-of-the-art approaches are being explored. For on-demand access, attested execution with trust bundles is being considered, along with federated identity and access management to overcome multi-factor authentication hurdles in urgent computing scenarios. Data movement solutions are being developed both for high-performance workflows within systems and for efficient cross-site transfers. Provenance tracking for AI models and object stores is becoming increasingly important, with potential solutions including the use of version control systems like Git or frameworks such as the HDF5 Common Metadata Framework~\cite{koomthanam2024common}, and the adoption of standards like RO-Crates~\cite{soiland2022packaging} for workflow provenance in major life science platforms. The FAIR4ML~\cite{fair4ml} initiative is proposing metadata schemas based on Schema.org and HuggingFace for capturing AI model descriptions. For data access policies, research is ongoing to develop methods that can guarantee secure access to object stores while complying with varying geographic regulations. The use of fabric-attached memory is being explored for HPC, although it presents new security challenges beyond traditional POSIX-style models.

\pp{Discussion}
Time-sensitive workflows are a key focus in scientific computing, driving efforts to refine definitions and research priorities. Emphasis is on distinguishing between ``time-critical" (\emph{real-time}) and ``time-sensitive" (\emph{workflow triggering}) processes, especially for near real-time control scenarios, such as device output to input parameters with time window constraints. The US Department of Energy (DOE) Office of Science's definition of time-sensitive workflows as ``low-latency workflows requiring real-time, or near-real-time, response across more than one facility or resource for timely decision-making and experiment steering" is recognized, emphasizing these workflows' multi-facility nature. A meta-analysis of 74 DOE Office of Science case studies found 45\% of workflows to be time-sensitive~\cite{dart2023esnet}.

Addressing time-sensitive workflow challenges involves key-value stores for smart applications, self-describing datasets for partial reads, and access pattern-based data migration. The expanding computing continuum (HPC, HTC, cloud, edge) necessitates advanced data movement and workflow management. This has led to interest in hierarchical WMSs, with cloud orchestrators coordinating specialized workflow managers for file-based, stream-focused, and other specialized tasks. Efforts focus on unifying WMSs to improve interoperability and reduce custom middleware. Robust provenance checking is emphasized, especially for long-term data and model tracking in regulated industries like fusion energy and life sciences.

Data access policies in Trusted Research Environments (TREs)~\cite{graham2023trust} in health data research are a key focus, with emphasis on controlled code movement and frameworks such as GA4GH passports~\cite{voisin2021ga4gh} and the Five Safes model~\cite{desai2016five}---they ensure data security and regulatory compliance across diverse geographic contexts. Trusted Execution Environments (TEEs), established by hardware extensions (e.g., Intel\textsuperscript{\textregistered} SGX, Intel\textsuperscript{\textregistered} TDX, or AMD\textsuperscript{\textregistered} SEV), are being explored for end-to-end distributed workflow confidentiality~\cite{Brescia:WiDE:2024,Brescia:ITADATA:2024}. Integrating AI and machine learning models into heavily regulated environments, such as power plant control systems, raises concerns about model construction, validation, and metadata management. 

The potential of serverless workflows in time-sensitive applications present both scalability advantages and data locality challenges. The scientific community has increased its efforts to enhance workflow portability across diverse computing environments, focusing on innovative solutions that leverage containerization technologies, virtual machines, and advanced scheduler integration techniques. These approaches aim to create more flexible, adaptable, and efficient WMSs capable of seamlessly operating across heterogeneous infrastructures. Initiatives like Autosubmit~\cite{manubens2016seamless}, which demonstrates portability across diverse systems including Fugaku, exemplify promising advances in workflow interoperability.

\recommendations{
   \begin{compactitem}
      \item Develop standardized definitions and metrics for time-sensitivity in workflows to facilitate better communication and comparison across different domains and applications.
       
      \item Invest in middleware solutions that can efficiently manage data movement and access across heterogeneous computing environments, focusing on both performance and security aspects.
       
      \item Establish best practices and frameworks for provenance tracking in AI and machine learning workflows, with emphasis on regulated environments and long-term data retention requirements.
       
      \item Unite workflow system developers, cloud service providers, and HPC centers to create portable solutions that can seamlessly operate across computing paradigms.
  \end{compactitem}
}



\section{Convergence of AI and HPC Workflows}
\label{sec:ai-workflows}

The integration of AI and HPC represents a paradigm shift in scientific and computational capabilities~\cite{badia2024integrating}. This convergence is multiplicative in its impact, enabling complex simulations and data-driven discoveries at unprecedented scales. AI-powered HPC workflows are drastically booting research efficiency across diverse scientific domains~\cite{ward2024employing, brewer2024ai}. While massive datasets generated by HPC simulations fuel the development and training of next-generation models implemented in AI systems, these AI systems, in turn, dynamically steer simulations for optimizing resource utilization and for maximizing scientific output. 

\pp{Emerging Paradigms and Challenges}
Large language models (LLMs) have become dominant, requiring exascale resources and motivating the development of workflows can perform zettaflops-scale computation. This shift has occurred alongside the continued growth of smaller-scale AI applications, including AI-driven robotics integrated with HPC workflows. The diversification of AI applications has expanded beyond proof-of-concept demonstrations to become integral components of larger frameworks and end-goals in themselves. However, persistent challenges remain. These include optimizing resource utilization, managing workflow performance and latency, and effectively routing data for inference tasks. Additionally, the community faces new hurdles in environment management, with the fast-paced AI ecosystem often outstripping traditional HPC build environments. Elasticity in compute and data capacity for AI training, along with reproducibility, reusability, and explainability of AI workflows on HPC systems, have emerged as critical concerns.

\pp{Recent Advances and Emerging Solutions}
To address these challenges, the community is enhancing existing workflows and developing decoupled data fabrics~\cite{suter2023escience}, with implementations based on ADIOS2 \cite{Adios2:2020} (e.g., INSTANT \cite{Instant:2023}), HDF5 \cite{HDF5:1999} (e.g., Wilkins \cite{Wilkins:2024}), or pure POSIX semantics (e.g., CAPIO \cite{CAPIO:2023}). This approach allows for greater flexibility in interchanging components to optimize workflows under varying conditions. AI startups have contributed to this trend by offering solutions that treat AI as distinct, interchangeable components within larger workflows. The focus has shifted towards creating modular, adaptable systems that can leverage the rapid advancements in AI techniques while maintaining the robustness required for HPC applications. Containerization in HPC is experiencing continued development, which promises to alleviate some of the environmental management issues. Cloud-service-oriented models are also being explored for their potential to mesh well with workflows, especially in multi-site scenarios. Despite the potential benefits, the adoption of cloud services in academic and national laboratory settings is tempered by concerns over cost structures, data governance, and the need to align with existing resource allocation models.

\pp{Discussion}
The integration of AI into HPC workflows has gained increased attention within the scientific community, centering on four key areas: cloud computing integration, provenance tracking, benchmark development, and AI-driven workflow optimization. While cloud services offer advantages such as elasticity and access to specialized hardware, concerns persist regarding cost models, data locality, and the blurring of boundaries between traditional HPC and cloud paradigms. The community recognizes the potential of cloud integration, particularly for specific workflow phases like post-processing or components requiring dynamic resource allocation, but acknowledges the need for careful consideration of implementation strategies in academic and national laboratory settings.

Provenance tracking in AI workflows has emerged as another key area of discussion. As AI models grow in complexity and are increasingly applied to critical decision-making processes, the ability to trace and explain model predictions has become paramount~\cite{10678731}. This need for trustworthy AI models, i.e., models that achieve a desirable balance between performance and principles like explainability, transparency, auditability, accountability, and reproducibility extends beyond the training phase to encompass the entire lifecycle of AI workflows, including the production and preparation of training data. The community faces the challenge of implementing comprehensive provenance tracking without introducing significant computational overhead, underscoring the need for efficient and standardized workflow provenance methodologies~\cite{cpe6544}.

The development of benchmarks for AI-HPC workflows and the potential of AI-driven workflow optimization have also garnered significant attention. While the community agrees on the importance of benchmarks for understanding performance bottlenecks and guiding hardware decisions, the diversity and rapid evolution of AI applications present challenges in establishing representative workflow patterns. Concurrently, the concept of ``AI for Workflows" is opening new avenues for automating the design and execution of complex HPC workflows. This paradigm shift raises intriguing questions about the optimal modularization of workflow components to facilitate AI-driven learning and composition.

\recommendations{
   \begin{compactitem}
       \item Develop frameworks for integrating cloud services with HPC workflows, addressing cost models, data lifecycle, and resource allocation to enable seamless hybrid cloud-HPC environments.
       
       \item Establish comprehensive provenance tracking for AI workflows, balancing thoroughness with efficiency to support explainability and reproducibility without sacrificing performance.
       
       \item Develop collaborative benchmarks representative of diverse AI-HPC workflow patterns across scientific domains, partnering domain experts with computer scientists.

       \item Invest in R\&D of AI-driven workflow optimization techniques, focusing on modular workflow designs that facilitate automated composition and adaptation of complex HPC workflows.

       \item Examine key requirements for parallel/distributed workflow patterns and models suited to large-scale AI/machine learning applications in science and engineering.
   \end{compactitem}
}


\section{Multi-Facility Workflows in Next-Generation Scientific Collaboration}
\label{sec:multi-facility}

Advanced instruments now produce vast amounts of complex scientific data. As individual experiment sites and computing facilities struggle to keep pace with unprecedented data rates and volumes, there is a growing recognition of the need for multi-facility (or federated) workflows~\cite{antypas2021enabling}. AI-enhanced multi-facility workflows integrate capabilities across HPC and experimental and observational facilities~\cite{ejarque2022enabling, badia2024integrating}, offering near real-time data analysis, improved experimental resilience, and better resource utilization.

\pp{Emerging Paradigms and Challenges}
The potential of multi-facility integrations has been recognized at the highest levels, with multiple finalists and winners of the Gordon Bell Special Prize showcasing composite workflows with downstream connections to user facilities. However, this emerging paradigm faces numerous challenges. These include multi-facility data movement, the need for curated large-scale data, and the persistent silos between science facilities. The heterogeneous hardware and software across HPC systems necessitate sophisticated, AI-enhanced middleware solutions for seamless workflow execution. Efficient data management across distributed environments requires robust transfer, storage, and security strategies. Additional hurdles include network constraints, batch-oriented HPC facilities with limited interactive capabilities, and project-based funding models that may not align with multi-facility collaborative efforts. Moreover, scheduling and resource allocation across independent facilities pose significant challenges. 

\pp{Recent Advances and Emerging Solutions}
The DOE's Integrated Research Infrastructure (IRI) program~\cite{brown2023integrated} leads these efforts with domain-specific pathfinder projects, while the field sees the emergence of both single-vendor ecosystems and cooperative system software stacks. The Superfacility~\cite{enders2020cross} project integrates experimental and computational resources across diverse scientific domains, enabling near real-time data analysis. In astronomy, the Dark Energy Camera (DECam) Alliance leverages AI algorithms to manage data across supercomputers and remote databases for identifying transient objects~\cite{tyler2022cross}. Despite these successes, there is a growing need for more fundamental, systemic solutions to support complex workflows across multiple facilities. Future research directions include exploring new scientific challenges suited to multi-facility resources, understanding AI workflow requirements, addressing AI at the edge, integrating quantum computing, and developing function-as-a-service capabilities.

\pp{Discussion}
Multi-facility workflows present a complex array of challenges and opportunities, mirroring the evolving landscape of large-scale scientific research. A key issue is the preparation of data for AI applications. Participants emphasized that much experimental data is ``dirty" -- unlabeled and noisy -- creating significant hurdles beyond conventional HPC and workflow concerns. This challenge is compounded by the complexities of cross-facility data transfer and user interaction, especially when navigating the varied internal policies and infrastructures of different facilities. The transition from individual user accounts to service accounts introduces additional layers of complexity in operational and security aspects. This shift underscores the urgent need for a standardized, robust authentication method that surpasses the security of current solutions like SSH, while maintaining ease of use, performance, and interoperability across diverse systems. Furthermore, workflow management across facilities presents its own set of unique challenges, including the non-trivial task of migrating workflows between machines with different architectures and policies (see Section~\ref{sec:heterogeneous}). Participants emphasized the critical importance of identifying and categorizing various interaction types and coupling modes within multi-facility hybrid workflows. A key step forward in addressing these challenges is enabling users to model distributed-by-design workflows. Such workflows would incorporate detailed descriptions of execution environments, required capabilities, and data movement restrictions directly into their representation, fostering more efficient and flexible multi-facility collaborations~\cite{HybridWorkflows:2023}.

Data movement across facilities remains a significant challenge. While some institutions use Globus~\cite{chard2017globus} for data transfer and workflow automation, others rely on alternatives like Rsync or custom solutions due to access limitations or security concerns. There is then a need for more universal and secure data movement solutions, especially for sensitive data that cannot leave certain jurisdictions (e.g., health and defense data).

To address these multifaceted challenges, participants proposed a strategic, multi-tiered approach. The immediate focus is on developing and showcasing demonstrable use cases, such as JAWS~\cite{tyler2022cross} and StreamFlow~\cite{StreamFlow}, supported by sustained funding for common solution development. Implementing multi-facility federated AI enables private, high-performance LLM training, inspired by projects like EuroHPC~\cite{eurohpc}. Concurrently, the development of multi-facility-aware workflow provenance standards was identified as a major step towards enhancing reproducibility, performance portability, and debugging capabilities across diverse environments~\cite{escience10254822}. Cross-Facility Federated Learning  (XFFL)~\cite{COLONNELLI2024XFFL} integrates workflow abstractions with Federated Learning, enabling cross-facility deep learning without large data transfers between sites. Multi-facility workflows face significant challenges, particularly in coordinating availability across multiple HPCs and managing job failures due to maintenance. These issues can invalidate federated training efforts. Developing robust fault tolerance mechanisms within new frameworks is crucial for ensuring successful execution of cross-facility collaborations and advancing scientific research~\cite{24:hipes:ft}.

Future goals include developing domain-agnostic workflows and distributed runtime systems, moving beyond centralized control/data plane orchestrators for greater flexibility and scalability. A key aspect of this strategy is the compilation of workflow models into distributed execution plans, leveraging multi-facility-aware, low-level intermediate representations. Looking at this direction, the SWIRL language~\cite{swirl} shifts the focus from high-level workflow languages, designed either for direct human interaction or for encoding complex, product-specific features, towards a low-level minimalistic representation of a workflow execution plan, making it more manageable for formalization methods and compiler toolchains. Last, discussions explored integrating AI and quantum computing into multi-facility workflows, addressing AI's unique requirements (e.g., AI-at-the-edge workflows) and quantum computing's need for classical fallback options due to current device unreliability~\cite{beck2024quantum}.

\recommendations{
   \begin{compactitem}
       \item Develop standardized cross-facility protocols for secure data movement, surpassing current solutions while ensuring interoperability and maintaining strict security standards.
       
       \item Implement distributed-by-design workflow modeling to create workflows inherently suited for multi-facility execution, incorporating environment details and constraints.
       
       \item Develop multi-facility-aware provenance standards and runtimes to compile workflows into facility-specific plans, improving reproducibility and performance across diverse environments.

       \item Explore and develop AI-ready data preparation solutions across facilities, addressing ``dirty" experimental data challenges and multi-facility AI workflow requirements.

       \item Investigate quantum computing integration in multi-facility workflows, developing classical simulation fallbacks for continuity and testing.
   \end{compactitem}
}


\section{Heterogeneous HPC Environments}
\label{sec:heterogeneous}

HPC is experiencing a radical shift towards increasingly heterogeneous environments. This transformation is clearly reflected in the latest Top500 list~\cite{top500}, where eight of the top ten systems heavily rely on GPUs for peak performance, and approximately 35\% of all listed machines incorporate hardware accelerators. The trend extends beyond traditional computing centers to encompass edge computing, creating a diverse computational continuum from edge devices to HPC systems and potentially cloud environments. As scientific workflows grow more sophisticated, often combining traditional HPC simulations with AI models and spanning multiple facilities, ensuring performance portability across these diverse systems has become a critical challenge.

\pp{Emerging Paradigms and Challenges}
Heterogeneous environments in HPC have become increasingly prevalent and complex in the past years. The introduction of exascale systems like Frontier at Oak Ridge National Laboratory and Aurora at Argonne National Laboratory, coupled with the pervasive influence of AI and LLMs, the emergence of specialized AI hardware, and the integration of dedicated AI partitions in large supercomputers, have fundamentally reshaped the HPC ecosystem. These advancements have intensified challenges in achieving performance portability across diverse hardware architectures, including CPUs, GPUs from various vendors, and specialized AI accelerators. Users now face a daunting array of tools and programming models, making it difficult to navigate the heterogeneous landscape efficiently. The integration of AI workflows with traditional HPC applications has created a convergence of two distinct computing paradigms, each with unique resource requirements and optimization strategies. Moreover, energy costs and system reliability have emerged as critical concerns in managing these intricate environments, adding further complexity to workflow design and execution.

\pp{Recent Advances and Emerging Solutions}
The HPC community is developing solutions for increasingly heterogeneous computing environments. A key focus is integrating workflow considerations into early HPC system design and deployment (e.g., including workflow benchmarks in the procurement process), creating more workflow-friendly architectures. Efforts to standardize resource management and workflow tools are ongoing, with projects like Flux~\cite{ahn2020flux} and libEnsemble~\cite{hudson2021libensemble} gaining prominence. These initiatives aim to simplify user experience across systems and reduce the learning curve for heterogeneous resources. Research into high-productivity, high-performance programming models is progressing, with languages like Julia and Python-based solutions offering accessible interfaces while leveraging low-level optimizations. There is also growing interest in harnessing AI and LLMs to assist with HPC programming and optimization tasks. The community is exploring regular virtual meetings and standardized workflow test suites to foster collaboration. These efforts aim to enhance portability, improve user training, and ensure effective utilization of heterogeneous HPC environments across scientific domains.

\pp{Discussion}
Heterogeneity in HPC environments encompasses a broad spectrum of contexts, including hardware diversity (CPUs, GPUs, and specialized accelerators), federated systems (multi-cluster, multi-site, facility-cloud integrations), varied applications and workflow tools, and data/storage heterogeneity. This multifaceted nature of heterogeneity underscores the complexity of challenges facing the HPC community and necessitates comprehensive solutions that address all aspects of heterogeneity in computing environments. 

The proliferation of HPC workflow tools like Parsl~\cite{babuji2019parsl}, RADICAL Cybertools~\cite{turilli2019middleware}, libEnsemble, PSI/J~\cite{hategan2023escience}, Merlin~\cite{peterson2022enabling}, SmartSim~\cite{partee2022using}, Dragon~\cite{radcliffe2023dragon}, AFCL~\cite{ristov2021afcl}, and
StreamFlow~\cite{StreamFlow}, alongside numerous homegrown solutions, underscores the need for better interoperability and common interfaces. These are key for developing and executing portable workflows across heterogeneous resources. There is then growing interest in developing a registry of execution environment information and standardizing resource description formats to improve resource discovery and allocation.

Scheduling and resource allocation in heterogeneous environments remain critical challenges. The complexities of both online (runtime) and offline (compile-time) scheduling underscore the need for sophisticated algorithms capable of handling the diverse constraints and optimization metrics inherent in heterogeneous systems~\cite{scheduling:parametric}. AI and LLMs show promise in addressing these challenges, particularly in areas like automated resource discovery, performance prediction, workflow optimization, and scheduling~\cite{scheduling:ai:gcn}.

To accelerate progress in managing and leveraging heterogeneous HPC environments, several initiatives are being considered. These include regular knowledge-sharing sessions, the development of comprehensive documentation on heterogeneous packages, and the creation of standardized workflow test suites~\cite{coleman2022pmbs, bard2023archetypes}. Such test suites could serve as benchmarks for evaluating different solutions and potentially aid in procurement decisions and system testing. These efforts aim to foster collaboration and innovation, ultimately enabling more efficient and effective scientific workflows across diverse computing resources.

\recommendations{
   \begin{compactitem}
        \item Create a central registry of execution environment information with standardized resource descriptions to enable discovery, allocation, and portability across heterogeneous HPC systems.
       
        \item Develop adaptive scheduling algorithms for efficient online and offline management in heterogeneous environments across diverse hardware, considering varied constraints and metrics.
                
        \item Create community-driven workflow benchmarks and test cases spanning diverse scientific domains and computational patterns to guide system evaluation, procurement, and optimization.
                
        \item Integrate AI and machine learning techniques, especially LLMs, into HPC workflow systems to automate resource discovery and optimization across heterogeneous computing environments.
    \end{compactitem}
}


\section{User Experience and Interfaces}
\label{sec:ux-ui}

As scientific workflows continue to evolve in complexity and scale, the importance of User Experience (UX)~\cite{paine2024} and intuitive interfaces has become increasingly critical. Well-designed UX can significantly impact the adoption, efficiency, and overall success of workflow systems, empowering researchers to focus on their scientific inquiries rather than grappling with technical intricacies. The interfaces (whether command line, graphical, API, or documentation) between scientists and workflow systems serve as a key bridge, determining how effectively researchers can harness the power of these tools.

\pp{Emerging Paradigms and Challenges}
UX in scientific workflows has gained increased recognition in recent years, with a broader understanding that UX encompasses more than just graphical user interfaces. The concept now includes the entire user journey, from product discovery to adoption, and extends to the policies and procedures governing services and facilities. This broadened perspective comes at a time of increasing complexity in the workflow landscape, characterized by a proliferation of systems and interfaces that reflect the growing heterogeneity of facilities and systems across national and global research infrastructures. This expanding ecosystem presents significant challenges for usability and interoperability among workflow systems. The integration of AI and machine learning into scientific workflows has also introduced new data curation challenges, affecting the adoption and integration of these tools. Current challenges include the need for seamless interactivity in heterogeneous environments, reducing barriers to entry for workflow adoption, and addressing the ``closed box" issue where workflow processes remain opaque to end-users, hindering troubleshooting and adaptability.

\pp{Recent Advances and Emerging Solutions}
Workflow system developers and UX researchers are actively working to simplify workflow usage, aiming to make it as intuitive as operating individual tools. Their efforts focus on easing the transition for scientists from manual tool chaining to the adoption of comprehensive workflow systems. Significant progress has been made in developing improved graphical representations of workflow steps and data flow, enhancing transparency and user comprehension. The scientific community is also adapting concepts from commercial software development, such as Design Systems, to the unique needs of scientific applications.  Initiatives like the Chan Zuckerberg Initiative's open-source science design system~\cite{eoss} and the STRUDEL project~\cite{strudel} are creating templates, code, and guidelines that enable teams to craft consistent and user-friendly interfaces across different scientific domains. On the other hand, the Jupyter Workflow \cite{JupyterWorkflow:2022} framework builds on the well-known Jupyter Notebook interface to let users design and run interactive large-scale distributed workflows with sequential equivalence guarantees. These efforts aim to improve the overall user experience of scientific workflows, from setup and configuration to execution and troubleshooting, ultimately fostering better collaboration between humans, machines, and research infrastructure systems.

\pp{Discussion}
UX in scientific workflows encompasses a wide spectrum of perspectives and implementations. Notable advancements in UX design have emerged, with some systems successfully integrating human participation within their graphical user interfaces for tasks such as quality control and experiment tuning. Industry-standard workflow platforms in bioinformatics, AI/ML, and cloud computing offer sophisticated UX features that serve as benchmarks for the field. However, replicating such interfaces in HPC environments presents unique challenges due to the inherent complexity of these ecosystems and their specialized technical and social requirements.

A central tension in workflow UX design lies between simplicity and complexity. There is a clear need to make workflows more accessible to newcomers, yet many researchers value the ability to engage with sophisticated tools as part of their learning and research processes. This dichotomy underscores the importance of developing flexible interfaces that can accommodate users with varying levels of expertise and diverse workflow requirements, allowing for both ease of use and depth of functionality~\cite{paine2021}.

The potential role of AI, particularly LLMs, in enhancing workflow UX is a subject of ongoing debate within the scientific community. While there is potential for AI to assist in generating workflow templates and improving documentation, concerns persist regarding the interpretability and maintainability of AI-generated workflows. These discussions highlight the need for thoughtful integration of AI technologies in scientific workflow systems, ensuring that they enhance rather than complicate the user experience.

Heterogeneity and portability remain critical challenges in workflow UX. The scientific community desires the flexibility to run workflows across different systems while still accessing specialized features of specific platforms. Successful approaches in addressing this challenge include developing systems that can automatically identify their execution environment and apply appropriate configurations. However, this raises fundamental questions about the optimal level of abstraction for workflows and how to balance the competing demands of portability and platform-specific optimization. 

\recommendations{
   \begin{compactitem}
       \item Formulate UX principles for scientific workflows, addressing common issues and guiding the design of intuitive, user-friendly interfaces (CLIs, APIs, GUIs) across various workflow~systems.
              
       \item Create tailored onboarding and training for new workflow system users, offering hands-on experience with real scientific problems and teaching core workflow management concepts.
              
       \item Develop adaptive interfaces for workflow systems that accommodate users of varying expertise, balancing simplicity for beginners with advanced features for experts.
       
       \item Explore responsible AI integration (LLMs) in workflow systems for assisted generation, documentation, and troubleshooting, while preserving human interpretability and maintainability.       
   \end{compactitem}
}

%

\section{FAIR Computational Workflows}
\label{sec:fair}

As scientific workflows become increasingly integral to data-intensive research across disciplines, ensuring their findability, accessibility, interoperability, and reusability (FAIR) has emerged as a critical challenge in the scientific community. The FAIR principles~\cite{wilkinson2016fair}, originally developed for research data, and later software~\cite{barker2022} are now being adapted to computational workflows to enhance their discoverability, reproducibility, and reuse. This shift is decisive as workflows embody not just data, but also scientific methods, analysis processes, and computational environments. By making workflows FAIR, researchers aim to improve collaboration, accelerate scientific discovery, and increase the overall efficiency of computational research.

\pp{Emerging Paradigms and Challenges}
The application of FAIR principles to computational workflows has emerged as a critical focus in the scientific community~\cite{goble2020}. As workflows become increasingly central to data-intensive research across disciplines, ensuring their FAIRness presents unique challenges. Unlike static datasets, workflows encompass both declarative descriptions and executable components, necessitating a nuanced approach to FAIR implementation. The complexity is further compounded by the diversity of workflow languages, execution environments, and domain-specific requirements. Emerging paradigms in workflow design, such as modular and composable workflows, raise questions about how to maintain FAIRness at various levels of granularity. Additionally, the integration of workflows with HPC environments and cloud infrastructures introduces new considerations for accessibility and portability. Balancing the need for standardization with the flexibility required for innovation remains a key challenge.

\pp{Recent Advances and Emerging Solutions}
Recent advances in making workflows FAIR have seen significant progress, with the development of FAIR principles specifically tailored to computational workflows~\cite{wilkinson2024FAIRwf}. These principles treat workflows as both data and software, addressing their unique dual nature. To support practical implementation, the community has developed infrastructure and tools such as DockStore~\cite{yuen2021dockstore} and WorkflowHub~\cite{goble2021implementing,gustafsson2024}, which facilitate the registration and sharing of workflow metadata, thereby enhancing findability and accessibility. Guidelines like ``Ten quick tips for building FAIR workflows''~\cite{de2023ten} provide researchers with actionable steps towards FAIRness. Efforts to improve interoperability between workflow management systems are ongoing, with a focus on rich metadata descriptions and standardized APIs. Emerging solutions also explore the potential of AI technologies in workflow programming and automated metadata generation. However, challenges remain in effectively embedding FAIR practices into workflow creation interfaces and extending the FAIR framework to encompass additional crucial aspects such as portability, reproducibility, and quality assurance \cite{kanwal2017investigating}. 

Within the Workflows Community Initiative, a working group has now established a set of principles for FAIR Computational Workflows~\cite{wilkinson2024FAIRwf}. These built on the FAIR principles for data and for research software~\cite{barker2022}, observed best practices, and considered workflows as an explainable composition of \emph{workflow components}, which themselves may be data sources, research software, or other workflows.

\pp{Discussion}
The application of the FAIR principles to computational workflows has sparked significant discussions within the scientific community, revealing several key challenges and potential directions for future development. A central topic is the definition and implementation of FAIR workflow components, which necessitates engagement with the broader ecosystem surrounding workflows, including dependencies, codes, and data. The challenge lies in striking a balance between comprehensive FAIRness and practicality, while avoiding the risk of overcomplicating the process by attempting to make every aspect FAIR.

Implementing FAIR principles in practice remains a significant hurdle in the workflow domain~\cite{wilkinson_f_2022}. Poor annotation of workflows and difficulties in sharing workflows across different languages are common issues. Proposed solutions include leveraging LLMs for workflow annotation, encouraging WMSs to adopt shared models for workflow description (such as the Bioschemas Computational Workflow profile~\cite{bcwp}), and promoting best practices among developers. The role of tools and services in encouraging FAIR practices is crucial, with efforts to engage tool developers in implementing FAIR-supporting features, such as the integration of Research Object Crate (RO-Crate)~\cite{soiland2022packaging} specifications in WMSs. This integration is demonstrated at~\cite{leo2024recording}, where six of these systems integrate the Workflow Run RO-Crate Profile.

There is growing interest in developing a FAIR workflow maturity model, drawing parallels with software maturity models and certification badges. Such a model could provide guidance for implementing FAIR principles and offer a framework for assessing the FAIRness of workflows. The development of FAIR metrics specifically tailored for workflows is seen as a key next step. Tools for FAIR assessment could automatically provide a FAIRness score or grade, presenting which principles are being evaluated within the tests~\cite{azevedo2024appraisal}. Tools like F-UJI~\cite{huber2021f} are being assessed for workflow applicability, with possible adaptations or new tools in development. The community is also exploring the relationship between FAIR metrics and maturity models, considering how they can complement each other in assessing and improving workflow FAIRness.

The application of FAIR principles to workflows containing export-controlled or confidential components presents unique challenges. This has highlighted the need for flexible approaches to FAIRness that can accommodate varying levels of openness and access restrictions. The concept of ``Inner FAIR'' within organizations has emerged as a potential solution, allowing for the application of FAIR principles even in contexts where full public sharing is not feasible. In areas in which sensitive data need to be managed, such as human genomics, users need to go through data access request procedures that are often manually managed by data access committees. This hinders the automation potential provided by computational workflow technologies. Initiatives such as the Data Use Ontology (DUO) \cite{lawsonDataUseOntology2021}, that allows for formally expressing the authorized uses for a dataset, could help to streamline tasks related to data discovery and retrieval in computational workflows that need to process controlled-access data.  Additionally, the community is grappling with the distinction between reusability and reproducibility in the context of workflows, recognizing that a workflow can be FAIR without necessarily being portable or reproducible across all environments.

\recommendations{
   \begin{compactitem}
       \item Promote standardized metadata schemas for workflows (Bioschemas ComputationalWorkflow profile, RO-Crate specifications) across WMSs to enhance interoperability and FAIR practices.
              
       \item Develop a FAIR workflow maturity model and metrics to guide FAIR implementation, offering standardized and automated assessment of workflow FAIRness, considering unique challenges of components and execution environments.
              
       \item Collaborate with tool developers to integrate FAIR-supporting features into workflow development tools, including automatic rich metadata generation, annotation support (leveraging AI technologies), and facilitation of workflow sharing across different languages and platforms.

       \item Develop guidelines and tools for implementing ``Inner FAIR" practices within organizations to address the challenges of applying FAIR principles to workflows with confidential or export-controlled components, enabling FAIR practices even when full public sharing is not possible.
   \end{compactitem}
}

\newpage
\cleardoublepage\phantomsection\addcontentsline{toc}{section}{References}
{\smaller 

}

\newpage
\cleardoublepage\phantomsection\addcontentsline{toc}{section}{Appendix A: Participants and Contributors}
\section*{Appendix A: Participants and Contributors}
\label{appx:contributors}

\vspace{-10pt}
\fontsize{10}{10}\selectfont
\begin{longtable}[!h]{llp{10.5cm}}
\toprule
\textbf{First Name} & \textbf{Last Name} & \textbf{Affiliation} \\
\midrule
    \rowcolor[HTML]{F2F2F2} 
    Abel & Souza & UMass Amherst \\ 
    Alberto & Mulone & University of Turin \\ 
    \rowcolor[HTML]{F2F2F2} 
    Amal & Gueroudji & Argonne National Laboratory \\ 
    Anderson Andrei & Da Silva & Hewlett Packard Enterprise Labs \\ 
    \rowcolor[HTML]{F2F2F2} 
    Andy & Gallo & GE Aerospace Research \\ 
    Barry & Sly-Delgado & University of Notre Dame \\ 
    \rowcolor[HTML]{F2F2F2} 
    Ben & Tovar & University of Notre Dame \\ 
    Billy & Rowell & Pacific Biosciences \\ 
    \rowcolor[HTML]{F2F2F2} 
    Brian & Etz & Oak Ridge National Laboratory \\ 
    Bruno & P. Kinoshita & Barcelona Supercomputing Center \\ 
    \rowcolor[HTML]{F2F2F2} 
    Carlo & Hamalainen & Standard Chartered \\ 
    Carole & Goble & The University of Manchester \\ 
    \rowcolor[HTML]{F2F2F2} 
    Chen & Li & UC Irvine \\ 
    Daniel & Balouek & Inria \\ 
    \rowcolor[HTML]{F2F2F2} 
    Daniel & de Oliveira & Universidade Federal Fluminense \\ 
    Daniel & Garijo & Universidad Politécnica de Madrid \\ 
    \rowcolor[HTML]{F2F2F2} 
    Daniel & Rosendo & Oak Ridge National Laboratory \\ 
    Daniela & Cassol & Lawrence Berkeley National Laboratory \\ 
    \rowcolor[HTML]{F2F2F2} 
    Deborah & Bard & National Energy Research Scientific Computing Center \\
    Dinindu & Senanayake & New Zealand eScience Infrastructure \\ 
    \rowcolor[HTML]{F2F2F2} 
    Domenico & Talia & Università della Calabria \\ 
    Doriana & Medi\'{c} & University of Turin \\ 
    \rowcolor[HTML]{F2F2F2} 
    Drew & Paine & Lawrence Berkeley National Laboratory \\ 
    Edite & Vartina & Riga Stradiņš University \\ 
    \rowcolor[HTML]{F2F2F2} 
    Eric & Vardar-Irrgang & Lawrence Livermore National Laboratory \\ 
    Fabian & Lehmann & Humboldt-Universität zu Berlin \\ 
    \rowcolor[HTML]{F2F2F2} 
    Fred & Suter & Oak Ridge National Laboratory \\ 
    Hector & Sanchez & Universidad de Puerto Rico \\ 
    \rowcolor[HTML]{F2F2F2} 
    Henri & Casanova & University of Hawaii at Manoa \\
    Iacopo & Colonnelli & University of Turin \\ 
    \rowcolor[HTML]{F2F2F2} 
    Ian & Foster & Argonne National Laboratory \\ 
    Jack & Marquez & University of Tennessee \\ 
    \rowcolor[HTML]{F2F2F2} 
    Jacob & Fosso Tande & North Carolina State University \\ 
    Jakob & Luettgau & Inria \\ 
    \rowcolor[HTML]{F2F2F2} 
    Jan & Janssen & Max Planck Institute for Sustainable Materials \\ 
    Jared & Coleman & Loyola Marymount University \\ 
    \rowcolor[HTML]{F2F2F2} 
    Jedrzej & Rybicki & Forschungszentrum Juelich \\ 
    Joe & Jordan & Uppsala University \\ 
    \rowcolor[HTML]{F2F2F2} 
    Johan & Gustafsson & Australia BioCommons \\ 
    John-Luke & Navarro & Argonne National Laboratory \\ 
    \rowcolor[HTML]{F2F2F2} 
    Jonathan & Bader & TU Berlin \\ 
    Jose & M. Fernandez & Barcelona Supercomputing Center \\ 
    \rowcolor[HTML]{F2F2F2} 
    Karan & Vahi & USC Information Sciences Institute \\ 
    Ketan & Maheshwari & Oak Ridge National Laboratory \\
    \rowcolor[HTML]{F2F2F2} 
    Kevin & Hunter Kesling & University of Chicago \\ 
    Khalid & Belhajjame & Université Paris-Dauphine \\
    \rowcolor[HTML]{F2F2F2} 
    Kin Wai & Ng & University of Tennessee \\ 
    Klaus & Noelp & University of Hagen \\ 
    \rowcolor[HTML]{F2F2F2} 
    Kyle & Chard & University of Chicago \\ 
    Laila & Los & University of Freiburg \\ 
    \rowcolor[HTML]{F2F2F2} 
    Lauren & Huet & University of Western Australia \\ 
    Lauritz & Thamsen & University of Glasgow \\ 
    \rowcolor[HTML]{F2F2F2} 
    Liliane & Kunstmann & UFRJ/COPPE \\ 
    Logan & Ward & Argonne National Laboratory \\ 
    \rowcolor[HTML]{F2F2F2} 
    Luiz & Gadelha & German Cancer Research Center \\
    Mahnoor & Zulfiqar & EMBL Heidelberg \\ 
    \rowcolor[HTML]{F2F2F2} 
    Marcos & Amaris & UFPA \\ 
    Merridee & Wouters & UNSW Sydney \\ 
    \rowcolor[HTML]{F2F2F2} 
    Michael & R. Crusoe & Common Workflow Language project \\ 
    Mikhail & Titov & Brookhaven National Laboratory \\ 
    \rowcolor[HTML]{F2F2F2} 
    Motohiko & Matsuda & RIKEN R-CCS \\ 
    Nathan & Tallent & Pacific Northwest National Laboratory \\ 
    \rowcolor[HTML]{F2F2F2} 
    Nishant & Saurabh & Utrecht University \\ 
    Nour & Elfaramawy & Humboldt University of Berlin \\ 
    \rowcolor[HTML]{F2F2F2} 
    Pascal & Elahi & Pawsey Supercomputing Research Centre \\ 
    Patricia & Grubel & Los Alamos National Laboratory \\ 
    \rowcolor[HTML]{F2F2F2} 
    Paul & Brunk & University of Georgia \\ 
    Paula & Iborra & Barcelona Supercomputing Center \\ 
    \rowcolor[HTML]{F2F2F2} 
    Peini & Liu & Barcelona Supercomputing Center \\ 
    Philipp & Gritsch & University of Innsbruck \\ 
    \rowcolor[HTML]{F2F2F2} 
    Qi & Yu & NIH \\ 
    Quentin & Guilloteau & University of Basel \\ 
    \rowcolor[HTML]{F2F2F2} 
    Rafael & Ferreira da Silva & Oak Ridge National Laboratory \\ 
    Rajat & Bhattarai & Tennessee Tech University \\ 
    \rowcolor[HTML]{F2F2F2} 
    Ra\"ul & Sirvent & Barcelona Supercomputing Center \\ 
    Renan & Souza & Oak Ridge National Laboratory \\ 
    \rowcolor[HTML]{F2F2F2} 
    Riccardo & Balin & Argonne National Lab \\ 
    Richard & Lupat & Peter MacCallum Cancer Centre \\ 
    \rowcolor[HTML]{F2F2F2} 
    Rolando & Hong Enriquez & Hewlett Packard Labs \\ 
    Rosa & Filgueira & EPCC, University of Edinburgh \\
    \rowcolor[HTML]{F2F2F2} 
    Ryan & Prout & Oak Ridge National Laboratory \\ 
    Sarah & Beecroft & Pawsey Supercomputing Research Centre \\ 
    \rowcolor[HTML]{F2F2F2} 
    Sashko & Ristov & University of Innsbruck \\ 
    Sean & Wilkinson & Oak Ridge National Laboratory \\ 
    \rowcolor[HTML]{F2F2F2} 
    Sehrish & Kanwal & University of Melbourne \\ 
    Shantenu & Jha & Princeton Plasma Physics Laboratory \\ 
    \rowcolor[HTML]{F2F2F2} 
    Shaun & de Witt & UK Atomic Energy Authority \\ 
    Shiva & Jahangiri & Santa Clara University \\ 
    \rowcolor[HTML]{F2F2F2} 
    Silvina & Caino-Lores & Inria \\ 
    Somayeh & Mohammadi & Mälardalen University \\ 
    \rowcolor[HTML]{F2F2F2} 
    Stefan & Robila & Montclair State University \\ 
    Stephen & Hudson & Argonne National Lab \\ 
    \rowcolor[HTML]{F2F2F2} 
    Stian & Soiland-Reyes & The University of Manchester \\ 
    Sumit Kumar & Saurav & C-DAC \\ 
    \rowcolor[HTML]{F2F2F2} 
    Taina & Coleman & San Diego Supercomputing Center \\ 
    Tanu & Malik & DePaul University \\ 
    \rowcolor[HTML]{F2F2F2} 
    Thomas & Fahringer & University of Innsbruck \\ 
    Tom & Gibbs & Nvidia \\ 
    \rowcolor[HTML]{F2F2F2} 
    Tom & Scogland & Lawrence Livermore National Laboratory \\ 
    Tyler & Skluzacek & Oak Ridge National Laboratory \\ 
    \rowcolor[HTML]{F2F2F2} 
    Ulf & leser & Humboldt-Universität zu Berlin \\ 
    Utz-Uwe & Haus & Hewlett Packard Enterprise Labs \\ 
    \rowcolor[HTML]{F2F2F2} 
    Wael & Elwasif & Oak Ridge National Laboratory \\ 
    Wes & Brewer & Oak Ridge National Laboratory \\ 
    \rowcolor[HTML]{F2F2F2} 
    Wesley & Ferreira & Universidade Federal Fluminense \\ 
    William F. & Godoy & Oak Ridge National Laboratory \\ 
    \rowcolor[HTML]{F2F2F2} 
    Woong & Shin & Oak Ridge National Laboratory \\ 
    Yadu & Babuji & University of Chicago \\ 
    \rowcolor[HTML]{F2F2F2} 
    Yiannis & Georgiou & Ryax Technologies \\ 
    Yuandou & Wang & University of Amsterdam \\ 
    \rowcolor[HTML]{F2F2F2} 
    Ziad & Al Bkhetan & University of Melbourne \\ 
    \bottomrule                                       
\end{longtable}

\fontsize{11}{11}\selectfont

\newpage
\cleardoublepage\phantomsection\addcontentsline{toc}{section}{Appendix B: Agenda}
\section*{Appendix B: Agenda}
\label{appx:agenda}

\vspace{-10pt}
\noindent
\fontsize{9}{10}\selectfont
\begin{longtable}{lp{14cm}}
    \toprule
    \multicolumn{2}{l}{\textbf{Day 1 -- September 3, 2024}} \\
    \midrule
    11:00-11:10am EDT & \makecell[l]{\textbf{Welcome and Introductions}} \\
    
    \rowcolor[HTML]{EEEEEE}
    11:10-11:20am EDT & \makecell[l]{\textbf{Frontiers in Scientific Workflows: Pervasive Integration with HPC}\\
    \emph{Rafael Ferreira da Silva (Oak Ridge National Laboratory)}} \\

    \rowcolor[HTML]{EEEEEE}
    11:20-11:50am EDT & \textbf{Lightning talks}
        \begin{compactitem}
            \item Topic 1: Time-Sensitive Workflows
            \item[] \emph{Utz-Uwe Haus (HPE), Shaun DeWitt (UKAEA)}
            \item Topic 2: Convergence of AI and HPC Workflows
            \item[] \emph{Shantenu Jha (PPPL), Logan Ward (ANL)}
            \item Topic 3: Multi-Facility Workflows in Next-Generation Scientific Collaboration
            \item[] \emph{Tom Gibbs (NVIDIA), Debbie Bard (LBNL)}
            \vspace{-10pt}
        \end{compactitem} \\
        
    11:50am-noon EDT & \emph{10min Break} \\
    
    \rowcolor[HTML]{EEEEEE}
    noon-1:30pm EST & \textbf{Breakout sessions} \\
    1:30-1:45pm EDT & \emph{15min Break} \\

    \rowcolor[HTML]{EEEEEE}
    1:45-2:00pm EDT & \textbf{Reports from breakout sessions} \\
    \bottomrule
\end{longtable}

\vspace{-10pt}
\noindent
\fontsize{9}{10}\selectfont
\begin{longtable}{lp{14cm}}
    \toprule
    \multicolumn{2}{l}{\textbf{Day 2 -- September 4, 2024}} \\
    \midrule
    11:00-11:10am EDT & \makecell[l]{\textbf{Summary and Highlights of Day 1} -- \emph{Stephen Hudson (ANL)}}\\
    
    \rowcolor[HTML]{EEEEEE}
    11:10-11:45am EDT & \textbf{Lightning talks}
        \begin{compactitem}
            \item Topic 4: Heterogeneous HPC Environments
            \item[] \emph{William Godoy (ORNL), Stephen Hudson (ANL)}
            \item Topic 5: User Experience and Interfaces
            \item[] \emph{Drew Paine (LBNL), Laila Los (UFreiburg)}
            \item Topic 6: FAIR Computational Workflows
            \item[] \emph{Carole Goble (UManchester), Sean Wilkinson (ORNL)}
            \vspace{-10pt}
        \end{compactitem} \\
        
    11:45am-noon EDT & \emph{15min Break} \\

    \rowcolor[HTML]{EEEEEE}
    noon-12:20pm EDT & \makecell[l]{\textbf{Invited talk:} AI-mediated Scientific Workflows --
    \emph{Ian T. Foster (ANL/UChicago)}} \\
    
    12:15-1:45pm EDT & \textbf{Breakout sessions} \\

    \rowcolor[HTML]{EEEEEE}
    1:45-2:00pm EDT & \textbf{Reports from breakout sessions} \\
    \bottomrule
\end{longtable}

\vspace{-10pt}
\noindent
\fontsize{9}{10}\selectfont
\begin{longtable}{lp{14cm}}
    \toprule
    \multicolumn{2}{l}{\textbf{Day 3 -- September 5, 2024} (Special Event for Asia/Oceania timezones)} \\
    \midrule
    12:30-12:35 AEST & \makecell[l]{\textbf{Welcome and Introductions}\\
    \emph{Sean Wilkinson (ORNL), Johan Gustafsson (Australian BioCommons)}} \\

    \rowcolor[HTML]{EEEEEE}
    12:35-12:45 AEST & \makecell[l]{\textbf{Context setting presentation for the WCI} \\
    \emph{Sean Wilkinson (ORNL)}} \\

    12:45-12:55 AEST & \makecell[l]{\textbf{Perspective on Bioinformatics Workflows in Australia} \\
    \emph{Johan Gustafsson (Australian BioCommons, University of Melbourne)}} \\
    
    \rowcolor[HTML]{EEEEEE}
    12:55-13:35 AEST & \textbf{Lightning talks}
        \begin{compactitem}
            \item Clinical Genomics Workflows
            \item[] \emph{Sehrish Kanwal (University of Melbourne)}
            \item Heterogeneous HPC Environments 
            \item[] \emph{Stephen Hudson (ANL)}
            \item Workflow Execution with Seqera Platform
            \item[] \emph{Ziad Al Bkhetan (Australian BioCommons)}
            \item Migrating Complex Workflows to the Exascale: Challenges for Radio Astronomy
            \item[] Pascal Elahi (Pawsey, Kensington WA)
            \vspace{-10pt}
        \end{compactitem} \\
        
    13:35-13:40 AEST & \emph{5min Break} \\

    \rowcolor[HTML]{EEEEEE}
    13:40-14:40 AEST & \textbf{Breakout sessions}
    \begin{compactitem}
            \item Multi-Facility Workflows
            \item What makes a Workflow Good?
            \item Heterogeneous HPC Environments
            \vspace{-10pt}
        \end{compactitem} \\

    14:40-15:00 AEST & \textbf{Reports from breakout sessions} \\
    \bottomrule
\end{longtable}

\includepdf[pages=-]{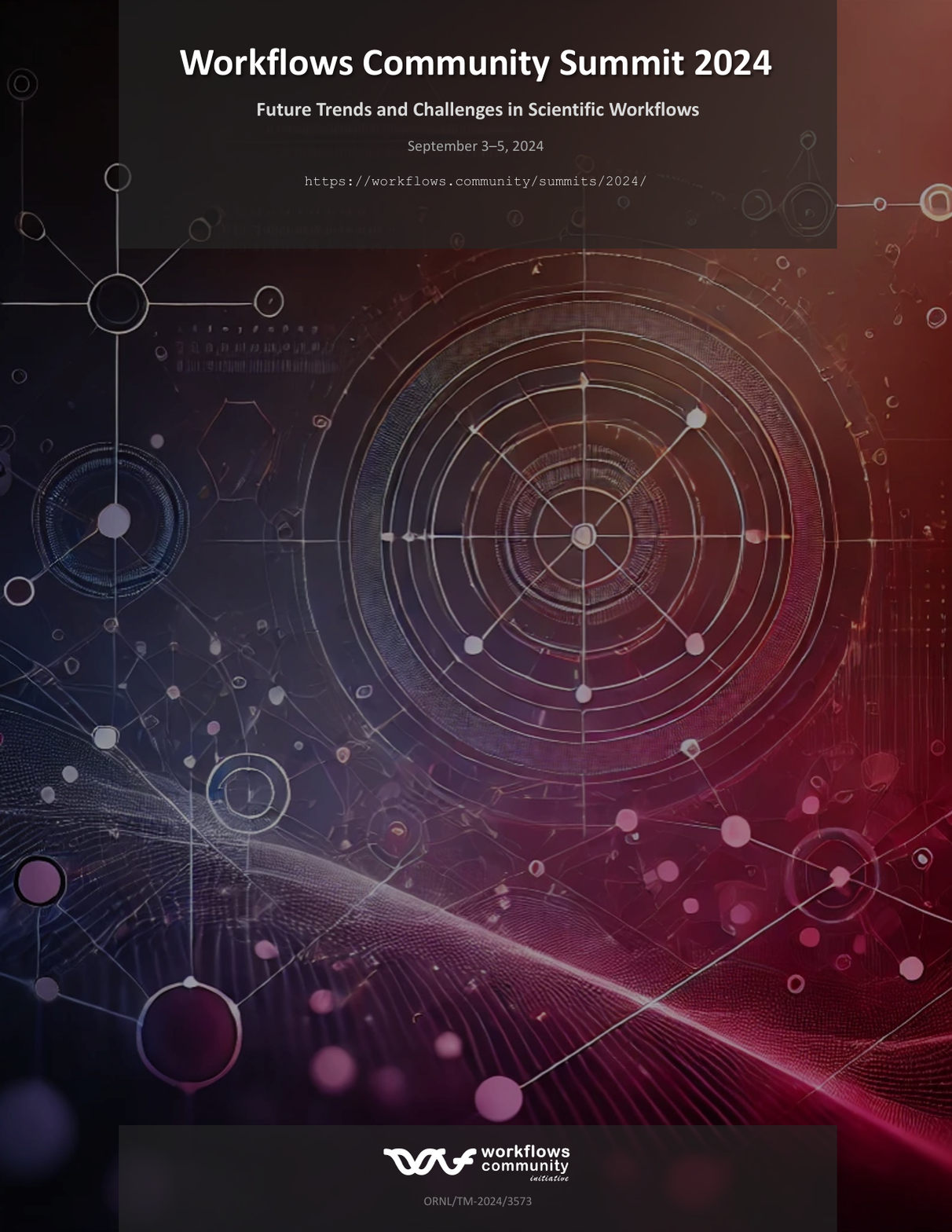}


\begin{thebibliography}{10}
\providecommand{\url}[1]{#1}
\csname url@samestyle\endcsname
\providecommand{\newblock}{\relax}
\providecommand{\bibinfo}[2]{#2}
\providecommand{\BIBentrySTDinterwordspacing}{\spaceskip=0pt\relax}
\providecommand{\BIBentryALTinterwordstretchfactor}{4}
\providecommand{\BIBentryALTinterwordspacing}{\spaceskip=\fontdimen2\font plus
\BIBentryALTinterwordstretchfactor\fontdimen3\font minus \fontdimen4\font\relax}
\providecommand{\BIBforeignlanguage}[2]{{%
\expandafter\ifx\csname l@#1\endcsname\relax
\typeout{** WARNING: IEEEtran.bst: No hyphenation pattern has been}%
\typeout{** loaded for the language `#1'. Using the pattern for}%
\typeout{** the default language instead.}%
\else
\language=\csname l@#1\endcsname
\fi
#2}}
\providecommand{\BIBdecl}{\relax}
\BIBdecl

\bibitem{ejarque2022enabling}
\BIBentryALTinterwordspacing
J.~Ejarque, R.~M. Badia, L.~Albertin, G.~Aloisio, E.~Baglione, Y.~Becerra, S.~Boschert, J.~R. Berlin, A.~D’Anca, D.~Elia \emph{et~al.}, ``{Enabling dynamic and intelligent workflows for HPC, data analytics, and AI convergence},'' \emph{Future generation computer systems}, vol. 134, 2022. [Online]. Available: \url{https://doi.org/10.1016/j.future.2022.04.014}
\BIBentrySTDinterwordspacing

\bibitem{brewer2024ai}
\BIBentryALTinterwordspacing
W.~Brewer, A.~Gainaru, F.~Suter, F.~Wang, M.~Emani, and S.~Jha, ``{AI-coupled HPC Workflow Applications, Middleware and Performance},'' \emph{arXiv preprint arXiv:2406.14315}, 2024. [Online]. Available: \url{https://doi.org/10.48550/arXiv.2406.14315}
\BIBentrySTDinterwordspacing

\bibitem{ferreiradasilva2024computer}
\BIBentryALTinterwordspacing
R.~Ferreira~da Silva, R.~M. Badia, D.~Bard, I.~T. Foster, S.~Jha, and F.~Suter, ``{Frontiers in Scientific Workflows: Pervasive Integration with HPC},'' \emph{IEEE Computer}, vol.~57, no.~8, 2024. [Online]. Available: \url{https://doi.org/10.1109/MC.2024.3401542}
\BIBentrySTDinterwordspacing

\bibitem{badia2024integrating}
\BIBentryALTinterwordspacing
R.~M. Badia, L.~Berti-Equille, R.~Ferreira~da Silva, and U.~Leser, ``{Integrating HPC, AI, and Workflows for Scientific Data Analysis},'' \emph{Dagstuhl Reports}, vol.~13, no.~8, 2024. [Online]. Available: \url{https://doi.org/10.4230/DagRep.13.8.129}
\BIBentrySTDinterwordspacing

\bibitem{antypas2021enabling}
\BIBentryALTinterwordspacing
K.~B. Antypas, D.~Bard, J.~P. Blaschke, R.~S. Canon, B.~Enders, M.~A. Shankar, S.~Somnath, D.~Stansberry, T.~D. Uram, and S.~R. Wilkinson, ``{Enabling Discovery Data Science Through Cross-Facility Workflows},'' in \emph{2021 IEEE International Conference on Big Data (Big Data)}.\hskip 1em plus 0.5em minus 0.4em\relax IEEE, 2021. [Online]. Available: \url{https://doi.org/10.1109/BigData52589.2021.9671421}
\BIBentrySTDinterwordspacing

\bibitem{wcs-2024}
``{Workflows Community Summit 2024},'' \url{https://workflows.community/summits/2024}, 2022.

\bibitem{wci}
``{Workflows Community Initiative},'' \url{https://workflows.community}, 2024.

\bibitem{francis_australian_2024}
\BIBentryALTinterwordspacing
R.~Francis and J.~H. Christiansen, ``Australian {BioCommons} {Strategic} {Plan} 2023 - 2028,'' 2024. [Online]. Available: \url{https://doi.org/10.5281/zenodo.13626350}
\BIBentrySTDinterwordspacing

\bibitem{ferreiradasilva2021works}
\BIBentryALTinterwordspacing
R.~Ferreira~da Silva, H.~Casanova, K.~Chard, I.~Altintas, R.~M. Badia, B.~Balis, T.~a. Coleman, F.~Coppens, F.~Di~Natale, B.~Enders, T.~Fahringer, R.~Filgueira \emph{et~al.}, ``{A Community Roadmap for Scientific Workflows Research and Development},'' in \emph{2021 IEEE Workshop on Workflows in Support of Large-Scale Science (WORKS)}, 2021. [Online]. Available: \url{https://doi.org/10.1109/WORKS54523.2021.00016}
\BIBentrySTDinterwordspacing

\bibitem{wcs2022}
\BIBentryALTinterwordspacing
R.~Ferreira~da Silva, R.~M. Badia, V.~Bala, D.~Bard, T.~Bremer, I.~Buckley, S.~Caino-Lores, K.~Chard, C.~Goble, S.~Jha, D.~S. Katz, D.~Laney, M.~Parashar, F.~Suter, N.~Tyler, T.~Uram, I.~Altintas \emph{et~al.}, ``{Workflows Community Summit 2022: A Roadmap Revolution},'' Oak Ridge National Laboratory, Tech. Rep. ORNL/TM-2023/2885, 2023. [Online]. Available: \url{https://doi.org/10.5281/zenodo.7750670}
\BIBentrySTDinterwordspacing

\bibitem{brown2022workflows}
\BIBentryALTinterwordspacing
N.~Brown, R.~Nash, G.~Gibb, E.~Belikov, A.~Podobas, W.~Der~Chien, S.~Markidis, M.~Flatken, and A.~Gerndt, ``Workflows to driving high-performance interactive supercomputing for urgent decision making,'' in \emph{International Conference on High Performance Computing}.\hskip 1em plus 0.5em minus 0.4em\relax Springer, 2022. [Online]. Available: \url{https://doi.org/10.1007/978-3-031-23220-6_16}
\BIBentrySTDinterwordspacing

\bibitem{brown2023integrated}
\BIBentryALTinterwordspacing
B.~L. Brown, W.~L. Miller \emph{et~al.}, ``{Integrated Research Infrastructure Architecture Blueprint Activity},'' Technical Report. US Department of Energy (USDOE), Washington, DC, Tech. Rep., 2023. [Online]. Available: \url{https://doi.org/10.2172/1984466}
\BIBentrySTDinterwordspacing

\bibitem{wedi2022destination}
\BIBentryALTinterwordspacing
N.~Wedi, P.~Bauer, I.~Sandu, J.~Hoffmann, S.~Sheridan, R.~Cereceda, T.~Quintino, D.~Thiemert, and T.~Geenen, ``{Destination Earth: High-Performance Computing for Weather and Climate},'' \emph{Computing in Science \& Engineering}, vol.~24, no.~6, 2022. [Online]. Available: \url{https://doi.org/10.1109/MCSE.2023.3260519}
\BIBentrySTDinterwordspacing

\bibitem{koomthanam2024common}
\BIBentryALTinterwordspacing
A.~J. Koomthanam, S.~Serebryakov, A.~Tripathy, G.~Nayak, M.~Foltin, and S.~Bhattacharya, ``{Common Metadata Framework: Integrated Framework for Trustworthy AI Pipelines},'' \emph{IEEE Internet Computing}, 2024. [Online]. Available: \url{https://doi.org/10.1109/MIC.2024.3377170}
\BIBentrySTDinterwordspacing

\bibitem{soiland2022packaging}
\BIBentryALTinterwordspacing
S.~Soiland-Reyes, P.~Sefton, M.~Crosas, L.~J. Castro, F.~Coppens, J.~M. Fern{\'a}ndez, D.~Garijo, B.~Gr{\"u}ning, M.~La~Rosa, S.~Leo \emph{et~al.}, ``{Packaging Research Artefacts with RO-Crate},'' \emph{Data Science}, vol.~5, no.~2, 2022. [Online]. Available: \url{https://doi.org/10.3233/DS-210053}
\BIBentrySTDinterwordspacing

\bibitem{fair4ml}
``{FAIR for Machine Learning (FAIR4ML) IG},'' \url{https://www.rd-alliance.org/groups/fair-machine-learning-fair4ml-ig}, 2024.

\bibitem{dart2023esnet}
\BIBentryALTinterwordspacing
E.~Dart \emph{et~al.}, ``{ESnet Requirements Review Program Through the IRI Lens: A Meta-Analysis of Workflow Patterns Across DOE Office of Science Programs},'' Lawrence Berkeley National Laboratory (LBNL), Berkeley, CA (United States), Tech. Rep., 2023. [Online]. Available: \url{https://doi.org/10.2172/2008205}
\BIBentrySTDinterwordspacing

\bibitem{graham2023trust}
\BIBentryALTinterwordspacing
M.~Graham, R.~Milne, P.~Fitzsimmons, and M.~Sheehan, ``{Trust and the Goldacre Review: Why Trusted Research Environments are not About Trust},'' \emph{Journal of Medical Ethics}, vol.~49, no.~10, 2023. [Online]. Available: \url{https://doi.org/10.1136/jme-2022-108435}
\BIBentrySTDinterwordspacing

\bibitem{voisin2021ga4gh}
\BIBentryALTinterwordspacing
C.~Voisin, M.~Linden, S.~O. Dyke, S.~R. Bowers, P.~Alper, M.~P. Barkley, D.~Bernick, J.~Chao, M.~Courtot, F.~Jeanson \emph{et~al.}, ``{GA4GH Passport Standard for Digital Identity and Access Permissions},'' \emph{Cell Genomics}, vol.~1, no.~2, 2021. [Online]. Available: \url{https://doi.org/10.1016/j.xgen.2021.100030}
\BIBentrySTDinterwordspacing

\bibitem{desai2016five}
\BIBentryALTinterwordspacing
T.~Desai, F.~Ritchie, R.~Welpton \emph{et~al.}, ``{Five Safes: Designing Data Access for Research},'' \emph{Economics Working Paper Series}, vol. 1601, p.~28, 2016. [Online]. Available: \url{https://doi.org/10.13140/RG.2.1.3661.1604}
\BIBentrySTDinterwordspacing

\bibitem{Brescia:WiDE:2024}
\BIBentryALTinterwordspacing
L.~Brescia and M.~Aldinucci, ``Secure generic remote workflow execution with {TEEs},'' in \emph{Proceedings of the 2nd Workshop on Workflows in Distributed Environments, WiDE 2024}.\hskip 1em plus 0.5em minus 0.4em\relax {ACM}, 2024. [Online]. Available: \url{https://doi.org/10.1145/3642978.3652834}
\BIBentrySTDinterwordspacing

\bibitem{Brescia:ITADATA:2024}
\BIBentryALTinterwordspacing
L.~Brescia, I.~Colonnelli, and M.~Aldinucci, ``Performance analysis on {DNA} alignment workload with {Intel SGX} multithreading,'' in \emph{Proceedings of the 2nd Special Track on Big Data and High-Performance Computing (BigHPC 2024)}, ser. {CEUR} Workshop Proceedings, vol. 3785.\hskip 1em plus 0.5em minus 0.4em\relax CEUR-WS.org, 2024. [Online]. Available: \url{https://ceur-ws.org/Vol-3606/invited77.pdf}
\BIBentrySTDinterwordspacing

\bibitem{manubens2016seamless}
D.~Manubens-Gil, J.~Vegas-Regidor, C.~Prodhomme, O.~Mula-Valls, and F.~J. Doblas-Reyes, ``Seamless management of ensemble climate prediction experiments on hpc platforms,'' in \emph{2016 International Conference on High Performance Computing \& Simulation (HPCS)}.\hskip 1em plus 0.5em minus 0.4em\relax IEEE, 2016.

\bibitem{ward2024employing}
\BIBentryALTinterwordspacing
L.~Ward, J.~G. Pauloski, V.~Hayot-Sasson, Y.~Babuji, A.~Brace, R.~Chard, K.~Chard, R.~Thakur, and I.~Foster, ``{Employing Artificial Intelligence to Steer Exascale Workflows with Colmena},'' \emph{arXiv preprint arXiv:2408.14434}, 2024. [Online]. Available: \url{https://doi.org/10.48550/arXiv.2408.14434}
\BIBentrySTDinterwordspacing

\bibitem{suter2023escience}
\BIBentryALTinterwordspacing
F.~Suter, R.~Ferreira~da Silva, A.~Gainaru, and S.~Klasky, ``{Driving Next-Generation Workflows from the Data Plane},'' in \emph{19th IEEE Conference on eScience}, 2023. [Online]. Available: \url{https://doi.org/10.1109/e-Science58273.2023.10254849}
\BIBentrySTDinterwordspacing

\bibitem{Adios2:2020}
\BIBentryALTinterwordspacing
W.~F. Godoy, N.~Podhorszki, R.~Wang, C.~Atkins, G.~Eisenhauer, J.~Gu, P.~E. Davis, J.~Choi, K.~Germaschewski, K.~A. Huck, A.~Huebl, M.~Kim \emph{et~al.}, ``{ADIOS} 2: The adaptable input output system. {A} framework for high-performance data management,'' \emph{SoftwareX}, vol.~12, 2020. [Online]. Available: \url{https://doi.org/10.1016/J.SOFTX.2020.100561}
\BIBentrySTDinterwordspacing

\bibitem{Instant:2023}
\BIBentryALTinterwordspacing
F.~Li and F.~Song, ``{INSTANT:} {A} runtime framework to orchestrate in-situ workflows,'' in \emph{Euro-Par 2023: Parallel Processing - 29th International Conference on Parallel and Distributed Computing, Limassol, Cyprus, August 28 - September 1, 2023, Proceedings}, vol. 14100.\hskip 1em plus 0.5em minus 0.4em\relax Springer, 2023. [Online]. Available: \url{https://doi.org/10.1007/978-3-031-39698-4\_14}
\BIBentrySTDinterwordspacing

\bibitem{HDF5:1999}
M.~Folk, A.~Cheng, and K.~Yates, ``{HDF5}: A file format and {I/O} library for high performance computing applications,'' in \emph{SC '99: Proc. of the {ACM/IEEE} Conf. on Supercomputing}, vol.~99, 1999.

\bibitem{Wilkins:2024}
\BIBentryALTinterwordspacing
O.~Yildiz, D.~Morozov, A.~Nigmetov, B.~Nicolae, and T.~Peterka, ``Wilkins: {HPC} in situ workflows made easy,'' \emph{CoRR}, vol. abs/2404.03591, 2024. [Online]. Available: \url{https://doi.org/10.48550/ARXIV.2404.03591}
\BIBentrySTDinterwordspacing

\bibitem{CAPIO:2023}
\BIBentryALTinterwordspacing
A.~R. Martinelli, M.~Torquati, M.~Aldinucci, I.~Colonnelli, and B.~Cantalupo, ``{CAPIO:} a middleware for transparent {I/O} streaming in data-intensive workflows,'' in \emph{30th {IEEE} International Conference on High Performance Computing, Data, and Analytics, HiPC 2023, Goa, India, December 18-21, 2023}.\hskip 1em plus 0.5em minus 0.4em\relax {IEEE}, 2023. [Online]. Available: \url{https://doi.org/10.1109/HIPC58850.2023.00031}
\BIBentrySTDinterwordspacing

\bibitem{10678731}
\BIBentryALTinterwordspacing
R.~Souza, S.~Caino-Lores, M.~Coletti, T.~J. Skluzacek, A.~Costan, F.~Suter, M.~Mattoso, and R.~F. Da~Silva, ``{Workflow Provenance in the Computing Continuum for Responsible, Trustworthy, and Energy-Efficient {AI}},'' in \emph{2024 IEEE 20th International Conference on e-Science (e-Science)}, 2024. [Online]. Available: \url{https://doi.org/10.1109/e-Science62913.2024.10678731}
\BIBentrySTDinterwordspacing

\bibitem{cpe6544}
\BIBentryALTinterwordspacing
R.~Souza, L.~G.~Azevedo, V.~Lourenço, E.~Soares, R.~Thiago, R.~Brandão, D.~Civitarese, E.~Vital~Brazil, M.~Moreno, P.~Valduriez, M.~Mattoso, R.~Cerqueira, and M.~A.~S.~Netto, ``Workflow provenance in the lifecycle of scientific machine learning,'' \emph{Concurrency and Computation: Practice and Experience}, vol. e6544, 2021. [Online]. Available: \url{https://doi.org/10.1002/cpe.6544}
\BIBentrySTDinterwordspacing

\bibitem{enders2020cross}
\BIBentryALTinterwordspacing
B.~Enders, D.~Bard, C.~Snavely, L.~Gerhardt, J.~Lee, B.~Totzke, K.~Antypas, S.~Byna, R.~Cheema, S.~Cholia \emph{et~al.}, ``{Cross-Facility Science with the Superfacility Project at LBNL},'' in \emph{2020 IEEE/ACM 2nd Annual Workshop on Extreme-scale Experiment-in-the-Loop Computing (XLOOP)}.\hskip 1em plus 0.5em minus 0.4em\relax IEEE, 2020. [Online]. Available: \url{https://doi.org/10.1109/XLOOP51963.2020.00006}
\BIBentrySTDinterwordspacing

\bibitem{tyler2022cross}
\BIBentryALTinterwordspacing
N.~Tyler, R.~Knop, D.~Bard, and P.~Nugent, ``{Cross-Facility Workflows: Case Studies with Active Experiments},'' in \emph{2022 IEEE/ACM Workshop on Workflows in Support of Large-Scale Science (WORKS)}.\hskip 1em plus 0.5em minus 0.4em\relax IEEE, 2022. [Online]. Available: \url{https://doi.org/10.1109/WORKS56498.2022.00014}
\BIBentrySTDinterwordspacing

\bibitem{HybridWorkflows:2023}
\BIBentryALTinterwordspacing
I.~Colonnelli, ``Workflow models for heterogeneous distributed systems,'' in \emph{Proceedings of the 2nd Italian Conference on Big Data and Data Science {(ITADATA} 2023)}, vol. 3606.\hskip 1em plus 0.5em minus 0.4em\relax CEUR-WS.org, 2023. [Online]. Available: \url{https://ceur-ws.org/Vol-3606/invited77.pdf}
\BIBentrySTDinterwordspacing

\bibitem{chard2017globus}
\BIBentryALTinterwordspacing
K.~Chard, I.~Foster, and S.~Tuecke, ``{Globus: Research Data Management as Service and Platform},'' in \emph{Proceedings of the Practice and Experience in Advanced Research Computing 2017 on Sustainability, Success and Impact}, 2017. [Online]. Available: \url{https://doi.org/10.1145/3093338.3093367}
\BIBentrySTDinterwordspacing

\bibitem{StreamFlow}
\BIBentryALTinterwordspacing
I.~Colonnelli, B.~Cantalupo, I.~Merelli, and M.~Aldinucci, ``{StreamFlow}: cross-breeding cloud with {HPC},'' \emph{{IEEE} {T}ransactions on {E}merging {T}opics in {C}omputing}, vol.~9, no.~4, 2021. [Online]. Available: \url{https://doi.org/10.1109/TETC.2020.3019202}
\BIBentrySTDinterwordspacing

\bibitem{eurohpc}
``{The European High Performance Computing Joint Undertaking (EuroHPC JU)},'' \url{https://eurohpc-ju.europa.eu/index_en}, 2024.

\bibitem{escience10254822}
\BIBentryALTinterwordspacing
R.~Souza, T.~J. Skluzacek, S.~R. Wilkinson, M.~Ziatdinov, and R.~Ferreira~da Silva, ``Towards lightweight data integration using multi-workflow provenance and data observability,'' in \emph{IEEE International Conference on e-Science}, 2023. [Online]. Available: \url{https://doi.org/10.1109/e-Science58273.2023.10254822}
\BIBentrySTDinterwordspacing

\bibitem{COLONNELLI2024XFFL}
\BIBentryALTinterwordspacing
I.~Colonnelli, R.~Birke, G.~Malenza, G.~Mittone, A.~Mulone, J.~Galjaard, L.~Y. Chen, S.~Bassini, G.~Scipione, J.~Martinovi\v{c}, V.~Vondr\'{a}k, and M.~Aldinucci, ``Cross-facility federated learning,'' \emph{Procedia Computer Science}, vol. 240, 2024. [Online]. Available: \url{https://doi.org/10.1016/j.procs.2024.07.003}
\BIBentrySTDinterwordspacing

\bibitem{24:hipes:ft}
A.~Mulone, D.~Medi\'{c}, and M.~Aldinucci, ``A fault tolerance mechanism for hybrid scientific workflows,'' in \emph{Proceedings of 1st workshop about {H}igh-{P}erformance e-{S}cience, HiPES, Madrid, Spain, 2024}, 2024.

\bibitem{swirl}
\BIBentryALTinterwordspacing
I.~Colonnelli, D.~Medi\'{c}, A.~Mulone, V.~Bono, L.~Padovani, and M.~Aldinucci, ``Introducing {SWIRL}: An intermediate representation language for scientific workflows,'' in \emph{Formal Methods. FM 2024}, vol. 14933.\hskip 1em plus 0.5em minus 0.4em\relax Springer Nature Switzerland, 2024. [Online]. Available: \url{https://doi.org/10.1007/978-3-031-71162-6\_12}
\BIBentrySTDinterwordspacing

\bibitem{beck2024quantum}
\BIBentryALTinterwordspacing
T.~Beck, A.~Baroni, R.~Bennink, G.~Buchs, E.~A. Coello~Perez, M.~Eisenbach, R.~Ferreira~da Silva, M.~Gopalakrishnan~Meena, K.~Gottiparthi, P.~Groszkowski, T.~S. Humble, R.~Landfield, K.~Maheshwari, S.~Oral, M.~A. Sandoval, A.~Shehata, I.-S. Suh, and C.~Zimmer, ``{Integrating Quantum Computing Resources into Scientific HPC Ecosystems},'' \emph{Future Generation Computer Systems}, vol. 161, pp. 11--25, 2024. [Online]. Available: \url{https://doi.org/10.1016/j.future.2024.06.058}
\BIBentrySTDinterwordspacing

\bibitem{top500}
``{TOP500},'' \url{https://top500.org}, 2024.

\bibitem{ahn2020flux}
\BIBentryALTinterwordspacing
D.~H. Ahn, N.~Bass, A.~Chu, J.~Garlick, M.~Grondona, S.~Herbein, H.~I. Ing{\'o}lfsson, J.~Koning, T.~Patki, T.~R. Scogland \emph{et~al.}, ``{Flux: Overcoming Scheduling Challenges for Exascale Workflows},'' \emph{Future Generation Computer Systems}, vol. 110, 2020. [Online]. Available: \url{https://doi.org/10.1109/WORKS.2018.00007}
\BIBentrySTDinterwordspacing

\bibitem{hudson2021libensemble}
\BIBentryALTinterwordspacing
S.~Hudson, J.~Larson, J.-L. Navarro, and S.~M. Wild, ``{libEnsemble: A Library to Coordinate the Concurrent Evaluation of Dynamic Ensembles of Calculations},'' \emph{IEEE Transactions on Parallel and Distributed Systems}, vol.~33, no.~4, 2021. [Online]. Available: \url{https://doi.org/10.1109/TPDS.2021.3082815}
\BIBentrySTDinterwordspacing

\bibitem{babuji2019parsl}
\BIBentryALTinterwordspacing
Y.~Babuji, A.~Woodard, Z.~Li, D.~S. Katz, B.~Clifford, R.~Kumar, L.~Lacinski, R.~Chard, J.~M. Wozniak, I.~Foster \emph{et~al.}, ``{Parsl: Pervasive Parallel Programming in Python},'' in \emph{Proceedings of the 28th International Symposium on High-Performance Parallel and Distributed Computing}, 2019. [Online]. Available: \url{https://doi.org/10.1145/3307681.3325400}
\BIBentrySTDinterwordspacing

\bibitem{turilli2019middleware}
\BIBentryALTinterwordspacing
M.~Turilli, V.~Balasubramanian, A.~Merzky, I.~Paraskevakos, and S.~Jha, ``{Middleware Building Blocks for Workflow Systems},'' \emph{Computing in Science \& Engineering}, vol.~21, no.~4, 2019. [Online]. Available: \url{https://doi.org/10.1109/MCSE.2019.2920048}
\BIBentrySTDinterwordspacing

\bibitem{hategan2023escience}
\BIBentryALTinterwordspacing
M.~Hategan-Marandiuc, A.~Merzky, N.~Collier, K.~Maheshwari, J.~Ozik, M.~Turilli, A.~Wilke, J.~M. Wozniak, K.~Chard, I.~Foster, R.~Ferreira~da Silva, S.~Jha, and D.~Laney, ``{PSI/J: A Portable Interface for Submitting, Monitoring, and Managing Jobs},'' in \emph{19th IEEE Conference on eScience}, 2023. [Online]. Available: \url{https://doi.org/10.1109/e-Science58273.2023.10254912}
\BIBentrySTDinterwordspacing

\bibitem{peterson2022enabling}
\BIBentryALTinterwordspacing
J.~L. Peterson, B.~Bay, J.~Koning, P.~Robinson, J.~Semler, J.~White, R.~Anirudh, K.~Athey, P.-T. Bremer, F.~Di~Natale \emph{et~al.}, ``{Enabling Machine Learning-Ready HPC Ensembles with Merlin},'' \emph{Future Generation Computer Systems}, vol. 131, 2022. [Online]. Available: \url{https://doi.org/10.1016/j.future.2022.01.024}
\BIBentrySTDinterwordspacing

\bibitem{partee2022using}
\BIBentryALTinterwordspacing
S.~Partee, M.~Ellis, A.~Rigazzi, A.~E. Shao, S.~Bachman, G.~Marques, and B.~Robbins, ``{Using Machine Learning at Scale in Numerical Simulations with SmartSim: An Application to Ocean Climate Modeling},'' \emph{Journal of Computational Science}, vol.~62, 2022. [Online]. Available: \url{https://doi.org/10.1016/j.jocs.2022.101707}
\BIBentrySTDinterwordspacing

\bibitem{radcliffe2023dragon}
\BIBentryALTinterwordspacing
N.~Radcliffe, K.~Lee, and P.~Mendygral, ``{Dragon Proxy Runtimes and Multi-system Workflows},'' in \emph{Proceedings of the SC'23 Workshops of The International Conference on High Performance Computing, Network, Storage, and Analysis}, 2023. [Online]. Available: \url{https://doi.org/10.1145/3624062.3624139}
\BIBentrySTDinterwordspacing

\bibitem{ristov2021afcl}
\BIBentryALTinterwordspacing
S.~Ristov, S.~Pedratscher, and T.~Fahringer, ``{AFCL: An Abstract Function Choreography Language for Serverless Workflow Specification},'' \emph{Future Generation Computer Systems}, vol. 114, 2021. [Online]. Available: \url{https://doi.org/10.1016/j.future.2020.08.012}
\BIBentrySTDinterwordspacing

\bibitem{scheduling:parametric}
\BIBentryALTinterwordspacing
J.~R. Coleman, R.~V. Agrawal, E.~Hirani, and B.~Krishnamachari, ``Parameterized task graph scheduling algorithm for comparing algorithmic components,'' \emph{CoRR}, vol. abs/2403.07112, 2024. [Online]. Available: \url{https://doi.org/10.48550/arXiv.2403.07112}
\BIBentrySTDinterwordspacing

\bibitem{scheduling:ai:gcn}
\BIBentryALTinterwordspacing
M.~Kiamari and B.~Krishnamachari, ``Gcnscheduler: scheduling distributed computing applications using graph convolutional networks,'' in \emph{Proceedings of the 1st International Workshop on Graph Neural Networking, GNNet 2022, Rome, Italy, 9 December 2022}.\hskip 1em plus 0.5em minus 0.4em\relax {ACM}, 2022. [Online]. Available: \url{https://doi.org/10.1145/3565473.3569185}
\BIBentrySTDinterwordspacing

\bibitem{coleman2022pmbs}
\BIBentryALTinterwordspacing
T.~Coleman, H.~Casanova, K.~Maheshwari, L.~Pottier, S.~R. Wilkinson, J.~Wozniak, F.~Suter, M.~Shankar, and R.~Ferreira~da Silva, ``{WfBench: Automated Generation of Scientific Workflow Benchmarks},'' in \emph{2022 IEEE/ACM International Workshop on Performance Modeling, Benchmarking and Simulation of High Performance Computer Systems (PMBS)}, 2022. [Online]. Available: \url{https://doi.org/10.1109/PMBS56514.2022.00014}
\BIBentrySTDinterwordspacing

\bibitem{bard2023archetypes}
D.~Bard, T.~Groves, B.~Cook, L.~Stephey, W.~Bhimji, S.~Farrell, B.~Austin, K.~Gott, S.~Canon, K.~Kallback-Rose, J.~Srinivasan, H.~A. Nam, and N.~J. Wright, ``{Workflow Archetypes White Paper},'' \url{https://www.nersc.gov/assets/NERSC-10/Workflows-Archetypes-White-Paper-v1.0.pdf}, National Energy Research Scientific Computing Center, Tech. Rep., 2023.

\bibitem{paine2024}
D.~Paine, ``Framing user experience (ux) across the scientific software lifecycle,'' \url{https://bssw.io/blog_posts/framing-user-experience-ux-across-the-scientific-software-lifecycle}, 2024.

\bibitem{eoss}
``{Essential Open Source Software for Science},'' \url{https://chanzuckerberg.com/eoss}, 2024.

\bibitem{strudel}
``{STRUDEL: Scientific sofTware Research for User experience, Design, Engagement, and Learning},'' \url{https://strudel.science/}, 2024.

\bibitem{JupyterWorkflow:2022}
\BIBentryALTinterwordspacing
I.~Colonnelli, M.~Aldinucci, B.~Cantalupo, L.~Padovani, S.~Rabellino, C.~Spampinato, R.~Morelli, R.~D. Carlo, N.~Magini, and C.~Cavazzoni, ``Distributed workflows with {Jupyter},'' \emph{Future Gener. Comput. Syst.}, vol. 128, 2022. [Online]. Available: \url{https://doi.org/10.1016/J.FUTURE.2021.10.007}
\BIBentrySTDinterwordspacing

\bibitem{paine2021}
D.~Paine, S.~Poon, and L.~Ramakrishnan, ``{Investigating User Experiences with Data Abstractions on High Performance Computing Systems},'' \url{https://doi.org/10.2172/1805039}, Lawrence Berkeley National Laboratory, Tech. Rep., 2021.

\bibitem{wilkinson2016fair}
\BIBentryALTinterwordspacing
M.~D. Wilkinson, M.~Dumontier \emph{et~al.}, ``{The {FAIR} Guiding Principles for Scientific Data Management and Stewardship},'' \emph{Scientific Data}, vol.~3, no.~1, 2016. [Online]. Available: \url{https://doi.org/10.1038/sdata.2016.18}
\BIBentrySTDinterwordspacing

\bibitem{barker2022}
\BIBentryALTinterwordspacing
M.~Barker, N.~P. Chue~Hong, D.~S. Katz, A.-L. Lamprecht, C.~Martinez-Ortiz, F.~Psomopoulos, J.~Harrow, L.~J. Castro, M.~Gruenpeter, P.~A. Martinez, and T.~Honeyman, ``Introducing the {FAIR} principles for research software,'' \emph{Scientific Data}, vol.~9, no.~1, 2022. [Online]. Available: \url{https://doi.org/10.1038/s41597-022-01710-x}
\BIBentrySTDinterwordspacing

\bibitem{goble2020}
\BIBentryALTinterwordspacing
C.~Goble, S.~Cohen-Boulakia, S.~Soiland-Reyes, D.~Garijo, Y.~Gil, M.~R. Crusoe, K.~Peters, and D.~Schober, ``{FAIR} {Computational Workflows},'' \emph{Data Intelligence}, vol.~2, no. 1–2, p. 108–121, 2020. [Online]. Available: \url{https://doi.org/10.1162/dint_a_00033}
\BIBentrySTDinterwordspacing

\bibitem{wilkinson2024FAIRwf}
\BIBentryALTinterwordspacing
S.~R. Wilkinson, M.~Aloqalaa, K.~Belhajjame, M.~R. Crusoe, B.~de~Paula~Kinoshita, L.~Gadelha, D.~Garijo, O.~J.~R. Gustafsson, N.~Juty, S.~Kanwal, F.~Z. Khan, J.~Köster, K.~P. von Gehlen, L.~Pouchard, R.~K. Rannow, S.~Soiland-Reyes, N.~Soranzo, S.~Sufi, Z.~Sun, B.~Vilne, M.~A. Wouters, D.~Yuen, and C.~Goble, ``Applying the {FAIR} principles to {Computational Workflows},'' \emph{arXiv}, 2024. [Online]. Available: \url{https://doi.org/10.48550/arXiv.2410.03490}
\BIBentrySTDinterwordspacing

\bibitem{yuen2021dockstore}
\BIBentryALTinterwordspacing
D.~Yuen, L.~Cabansay, A.~Duncan, G.~Luu, G.~Hogue, C.~Overbeck, N.~Perez, W.~Shands, D.~Steinberg, C.~Reid \emph{et~al.}, ``{The Dockstore: Enhancing a Community Platform for Sharing Reproducible and Accessible Computational Protocols},'' \emph{Nucleic Acids Research}, vol.~49, no.~W1, 2021. [Online]. Available: \url{https://doi.org/10.1093/nar/gkab346}
\BIBentrySTDinterwordspacing

\bibitem{goble2021implementing}
\BIBentryALTinterwordspacing
C.~Goble, S.~Soiland-Reyes, F.~Bacall, S.~Owen, A.~Williams, I.~Eguinoa, B.~Droesbeke, S.~Leo, L.~Pireddu, L.~Rodr{\'\i}guez-Navas \emph{et~al.}, ``{Implementing FAIR Digital Objects in the EOSC-Life Workflow Collaboratory},'' \emph{Zenodo}, 2021. [Online]. Available: \url{https://doi.org/10.5281/zenodo.4605654}
\BIBentrySTDinterwordspacing

\bibitem{gustafsson2024}
\BIBentryALTinterwordspacing
O.~J.~R. Gustafsson, S.~R. Wilkinson, F.~Bacall, S.~Soiland-Reyes, S.~Leo, L.~Pireddu, S.~Owen, N.~Juty, J.~M. Fernández, T.~Brown, H.~Ménager, B.~Grüning, S.~Capella-Gutierrez, F.~Coppens, and C.~Goble, ``{WorkflowHub}: a registry for computational workflows,'' \emph{arXiv}, 2024. [Online]. Available: \url{https://doi.org/10.48550/arXiv.2410.06941}
\BIBentrySTDinterwordspacing

\bibitem{de2023ten}
\BIBentryALTinterwordspacing
C.~de~Visser, L.~F. Johansson, P.~Kulkarni, H.~Mei, P.~Neerincx, K.~Joeri van~der Velde, P.~Horvatovich, A.~J. van Gool, M.~A. Swertz, P.~Hoen \emph{et~al.}, ``{Ten Quick Tips for Building FAIR Workflows},'' \emph{PLoS Computational Biology}, vol.~19, no.~9, 2023. [Online]. Available: \url{https://doi.org/10.1371/journal.pcbi.1011369}
\BIBentrySTDinterwordspacing

\bibitem{kanwal2017investigating}
S.~Kanwal, F.~Z. Khan, A.~Lonie, and R.~O. Sinnott, ``Investigating reproducibility and tracking provenance--a genomic workflow case study,'' \emph{BMC bioinformatics}, vol.~18, 2017.

\bibitem{wilkinson_f_2022}
\BIBentryALTinterwordspacing
S.~R. Wilkinson, G.~Eisenhauer, A.~J. Kapadia, K.~Knight, J.~Logan, P.~Widener, and M.~Wolf, ``F*** workflows: when parts of {FAIR} are missing,'' in \emph{2022 {IEEE} 18th {International} {Conference} on e-{Science} (e-{Science})}.\hskip 1em plus 0.5em minus 0.4em\relax IEEE, 2022. [Online]. Available: \url{https://doi.org/10.1109/eScience55777.2022.00090}
\BIBentrySTDinterwordspacing

\bibitem{bcwp}
``{Bioschemas ComputationalWorkflow Profile},'' \url{https://bioschemas.org/profiles/ComputationalWorkflow/1.0-RELEASE}, 2021.

\bibitem{leo2024recording}
\BIBentryALTinterwordspacing
S.~Leo, M.~R. Crusoe, L.~Rodr{\'\i}guez-Navas, R.~Sirvent, A.~Kanitz, P.~De~Geest, R.~Wittner, L.~Pireddu, D.~Garijo, J.~M. Fern{\'a}ndez \emph{et~al.}, ``{Recording provenance of workflow runs with RO-Crate},'' \emph{PLoS one}, vol.~19, no.~9, 2024. [Online]. Available: \url{https://doi.org/10.1371/journal.pone.0309210}
\BIBentrySTDinterwordspacing

\bibitem{azevedo2024appraisal}
\BIBentryALTinterwordspacing
L.~G. Azevedo, G.~Banaggia, J.~Tesolin, and R.~Cerqueira, ``An appraisal of automated tools for fairness evaluation,'' in \emph{4th Workshop on Metadata and Research Objects Management for Linked Open Science (DaMaLOS 2024)}, 2024. [Online]. Available: \url{https://repository.publisso.de/resource/frl:6483276}
\BIBentrySTDinterwordspacing

\bibitem{huber2021f}
\BIBentryALTinterwordspacing
R.~Huber and A.~Devaraju, ``{F-UJI: an Automated Tool for the Assessment and Improvement of the FAIRness of Research Data},'' in \emph{EGU General Assembly Conference Abstracts}, 2021. [Online]. Available: \url{https://doi.org/10.5194/egusphere-egu21-15922}
\BIBentrySTDinterwordspacing

\bibitem{lawsonDataUseOntology2021}
\BIBentryALTinterwordspacing
J.~Lawson, M.~N. Cabili, G.~Kerry, T.~Boughtwood, A.~Thorogood, P.~Alper, S.~R. Bowers, R.~R. Boyles \emph{et~al.}, ``{The {Data Use Ontology} to Streamline Responsible Access to Human Biomedical Datasets},'' \emph{Cell Genomics}, vol.~1, no.~2, 2021. [Online]. Available: \url{https://doi.org/10.1016/j.xgen.2021.100028}
\BIBentrySTDinterwordspacing

\end{thebibliography}
\end{document}